\documentclass[showpacs,prb,floatfix,superscriptaddress,twocolumn,amsmath]{revtex4-1}
\usepackage{graphics,epsfig,amsfonts,amssymb,amsmath,ulem,color}
\def\ra{\rightarrow}

\def\bk{{\bf k}}

\def\bq{{\bf q}}

\def\DD{{\cal{D}}}
\def\GG{{\cal{G}}}

\def\VV{{\cal{V}}}

\def\a{\alpha}
\def\b{\beta}
\def\e{\varepsilon}
\def\d{\delta}

\def\m{\mu}
\def\l{\lambda}

\def\t{\tau}

\def\s{\sigma}
\def\o{\omega}
\def\G{\Gamma}
\def\D{\Delta}
\def\O{\Omega}

\newcommand{\be}{\begin{equation}}
\newcommand{\ee}{\end{equation}}
\newcommand{\bea}{\begin{eqnarray}}
\newcommand{\eea}{\end{eqnarray}}
\def\nn{\nonumber}
\def\lb{\label}
\def\pref#1{(\ref{#1})}

\begin{document}
\title{Spin-orbital interplay and topology in the nematic phase of iron pnictides}

\author{Laura Fanfarillo}
\affiliation{Instituto de Ciencia de Materiales de Madrid, ICMM-CSIC, Cantoblanco, E-28049 Madrid (Spain).}
\author{Alberto Cortijo}
\affiliation{Instituto de Ciencia de Materiales de Madrid, ICMM-CSIC, Cantoblanco, E-28049 Madrid (Spain).}
\author{Bel\'en Valenzuela}
\affiliation{Instituto de Ciencia de Materiales de Madrid, ICMM-CSIC, Cantoblanco, E-28049 Madrid (Spain).}
\email{belenv@icmm.csic.es}
\date{26 May 2015}
\begin{abstract}
The origin of the nematic state is an important puzzle to be solved in iron 
pnictides. Iron superconductors are multiorbital systems and these orbitals play 
an important role at low energy. The singular $C_4$ symmetry of $d_{zx}$ and 
$d_{yz}$ orbitals has a profound influence at the Fermi surface since the 
$\Gamma$ pocket has vortex structure in the orbital space and the X/Y electron 
pockets have $yz$/$zx$ components respectively. We propose a low energy theory 
for the spin--nematic model derived from a multiorbital Hamiltonian. In the 
standard spin--nematic scenario the ellipticity of the electron pockets is a 
necessary condition for nematicity. In the present model nematicity is 
essentially due to the singular $C_4$ symmetry of $yz$ and $zx$ orbitals. By 
analyzing the ($\pi, 0$) spin susceptibility in the nematic phase we find 
spontaneous generation of orbital splitting extending previous calculations in 
the magnetic phase. We also find that the ($\pi, 0$) spin susceptibility has an 
intrinsic anisotropic momentum dependence due to the non trivial topology of the 
$\Gamma$ pocket. 
\end{abstract}

\maketitle

\section{INTRODUCTION}
Most iron pnictides parent compounds present $(\pi,0)$ antiferromagnetism (AF) 
in the 1--Fe Brillouin zone (BZ) and become superconductors upon doping or 
pressure. A structural transition takes place before or simultaneously to the AF 
one. These two transitions enclose the nematic phase characterized by many 
experimental probes \cite{mazin10, chuscience2010, fisher2011, yingprl11, 
chuscience2012, fisherprl14, jiangprl13, Fernandesprl10, meingastprl04, 
degiorgi10, uchida2011, degiorgi2012, nakajimaprl12, 
laura-greenprb12,prozorovnatcomm13, degiorgiprb14, gallaisprl13, sciencedavis10, 
davisnatphys2013, rosenthalnatphys14, matsudanat12, pengchengdaiprb11, 
dhitalprl12, pengchengdaiprb13, ibukaphysC14, luscience14}. Signatures of 
ferro--orbital ordering (OO) in the magnetic and nematic phase are found in 
ARPES \cite{shenpnas11, shennjp12} and X--ray absorption spectroscopy 
\cite{kim13}. The origin of the nematic phase is widely debated in the 
literature. There is a controversy on whether nematicity is mostly intrinsic or 
mostly due to impurity scattering \cite{uchidaprl13, davisnatphys2013, 
pengchengdaiprb13, andersenprl13, fisherprl14, hirschfeldarXiv14}. In the 
intrinsic scenario it has been discussed in the context of lattice, magnetic and 
orbital degrees of freedom (d.o.f) \cite{kivelson08, sachdevprb08, leeyinku09, 
nosotrasprl10-2, yinpickettprb10, lvphillipsprb10, chen_devereaux10, 
laadcracoprb11, canoprb10, canoprb11,canoprb12, schmalianprb12, 
Fernandesreview12, liangprl13, liangarXiv14, ipaulprb14, fernandesnatphys14, 
hirschfeldnatphys14, luscience14, buchner_nature15}. Different experiments suggest an 
electronic origin \cite{chuscience2010, shenpnas11, yingprl11, chuscience2012, 
matsudanat12, gallaisprl13,rosenthalnatphys14} but due to the spin--orbital 
interplay it is difficult to pinpoint between the two. 

First principle calculations and mean--field (MF) approaches in multiorbital 
Hamiltonians have found OO only inside the $(\pi,0)$ AF phase \cite{leeyinku09, 
yu09, nosotrasprl10, nosotrasprl10-2, daghofer10, yinpickettprb10, yin11}. 
Beyond MF, Random Phase Approximation \cite{graser09, hinkovprb10} or Dynamical 
Mean Field Theory \cite{kotliarprl11} calculations provide a good description of 
the spin dynamics in different compounds. However, the complexity of such 
calculations makes difficult to sort out the essential physics. On the other 
hand, the Landau approach is very useful to understand the interplay between the 
structural and magnetic d.o.f and to calculate response functions 
\cite{lorenzanaprl08, Fernandesprl10, chubukovprb10, canoprb12, lorenzanaprb11, 
brydonprb11, schmalianprb12, Fernandesreview12, fernandesnatphys14}, in 
particular in the context of the spin--nematic (S--N) theory.  Within this 
scenario we have to consider two broken symmetries in the $(\pi,0)$ AF phase: 
the $O(3)$ spin--rotational symmetry and the $Z_2$ S--N symmetry between the 
$(\pi,0)$ and $(0,\pi)$ AF state. The discrete $Z_2$ S--N symmetry can appear at 
higher temperature than the magnetic transition. In the work by Fernandes et 
al.~[\onlinecite{schmalianprb12}] the S--N phase is derived from a microscopic 
Hamiltonian with hole (h\,--) and electron (e\,--) pockets without structure in the 
orbital d.o.f, in the following named orbital--less effective approach. In this 
approach the nematic order parameter (OP) crucially depends on the ellipticity 
of the e\,--\,pockets vanishing for circular electron Fermi surfaces (FSs) 
\cite{schmalianprb12, chubukovprb10}. Ellipticity arguments have also been used 
to explain experiments \cite{schmalianprb12, prozorovnatcomm13,chubukovprb10, 
mcqueeneyprb12, Fernandesreview12}. In spite of its simplicity this model helps 
understand the interplay between the structural, nematic and magnetic 
transitions \cite{schmalianprb12}. However, the absence of microscopic 
information about the orbital d.o.f and the lack of connection between this 
approach and multiorbital Hamiltonians leaves several important questions open.

Mostly three iron orbitals contribute to the FS of iron superconductors, $yz$ 
and $zx$ for the $\Gamma$ pockets and $xy$, $yz$/$zx$ for the the X/Y pocket. 
The particular arrangement of the $yz$ and $zx$ orbitals arises because under a 
$\pi/2$ rotation the two orbitals transform as $|xz\rangle \rightarrow 
|yz\rangle$ and $|yz\rangle \rightarrow -|xz\rangle$ \cite{scalapinoarXiv08}. 
Important features of this singular $C_4$ symmetry are (i) the $yz/zx$ orbital 
content in the X/Y pocket and (ii) a non trivial topology in the $\Gamma$ pocket 
with vorticity equal 2 in the non--magnetic phase. Due to this topology a nodal 
spin density wave (SDW) with two Dirac cones is formed in the magnetic 
phase \cite{dunghailee09, lautimmprb13}. It is essential then to find out the 
consequences of this low energy physics in the S--N scenario.

We propose an effective action for the magnetic instability 
derived from a multiorbital Hamiltonian. The Landau coefficients depend on the 
orbital content, Hubbard and Hund's coupling. The orbital d.o.f. changes the 
S--N scenario in an essential way: (i) the $yz/zx$ orbital content of the X/Y 
e\,--\,pockets gives rise to a finite nematic coupling regardless the value of 
the ellipticity of the e\,--\,pockets.  Analyzing the ($\pi,0$) magnetic 
susceptibility in the nematic phase we find that: (ii) the  $zx$, $yz$ 
degeneracy is spontaneously broken without considering a small crystal field or 
interactions between the e\,--\,pockets in the original Hamiltonian and (iii) 
the spin fluctuation have anisotropic momentum dependence in the $x$, $y$ 
directions due to the spin interaction connecting the $\Gamma$ pocket with 
vorticity two with the topologically trivial $X$ e--pocket. The results obtained 
are robust for any number of orbitals since they are based in symmetry and 
topological arguments. To obtain analytical results, we analyze the continuum 
limit of the $d_{yz}$--\,$d_{zx}$ orbital model that displays the fundamental 
$C_4$ orbital symmetry. Within this approximation, we also consider the 
ellipticity of the e-pockets perturbatively to compare our results with previous 
theoretical expectations. We find that the vortex of the $\Gamma$ pocket gives 
an anisotropic $\bq$--dependence of the $(\pi,0)$ magnetic susceptibility 
already at zero order and also affects the contribution proportional to the 
ellipticity parameter. The calculation of the nematic coupling confirms that, in 
the limit of validity of the approximation, the dominant contribution is the one 
coming from the orbital symmetry.

\section{EFFECTIVE ACTION FOR THE MULTIORBITAL SYSTEM}  
We consider a multiorbital Hamiltonian for the FeAs layer including a 
tight--binding (TB) term plus local interactions as described in [\onlinecite{nosotrasprl10}]. 
By considering rotational invariance, interactions 
can be expressed in terms of two coefficients: the intra--orbital Hubbard $U$ and 
the Hund coupling $J_H$ \cite{castellani78}. The TB term can be diagonalized 
$c_{\eta \bk \alpha}^\dagger= \sum_n a_{\eta n}^* (\bk) d^\dagger_{n \bk 
\alpha}$ with $a_{\eta n}({\bf k})$ the rotation matrix element between the 
orbital $\eta$ and the band basis $n$. Since we are interested in the low energy 
physics we will restrict to energies and momenta close to the FS taking into 
account the h\,--\,pocket at $\Gamma$ and the $X$ and $Y$ e\,--\,pockets. For simplicity 
we do not consider the third pocket found at the $M$ point in the FS since it is 
parameter sensitive and it is not usually taken into account in the S--N 
scenario. Therefore we have $c_{\eta \bk \a }^\dagger=\sum_m a_{\eta 
m}^*(\bk)d^\dagger_{m \bk \a}$ with $m=\Gamma,X,Y$ pockets and ${\bf k}$ 
restricted to be close to the FS. Following [\onlinecite{schmalianprb12}], we will 
consider only the spin channel of the interaction of the Hamiltonian and we 
restrict to the spin excitations with momenta near ${\bf Q}_1=(\pi,0)$ and ${\bf 
Q}_2=(0,\pi)$. The interaction Hamiltonian is given by
\begin{equation}
H_{\text{int}}=-\frac{1}{2}\sum_{\bf q}\sum_{\eta_1\eta_2}\sum_{l=X,Y} U^{\text{spin}}_{\eta_1\eta_2} 
\vec{ S}_{\eta_1l}({\bf q})\cdot\vec{ S}_{\eta_2l}({-\bf q}).
\label{eq:interaction}
\end{equation}
$U^{\text{spin}}_{\eta_1\eta_2}=\frac{8}{3}U\delta_{\eta_1\eta_2} + 4J_H(1-\delta_{\eta_1\eta_2})$
is a matrix in the orbital space. 
$\vec{ S}_{\eta l}(\bq)=\sum_\bk w_{\G l}^\eta(\bk,\bk + \bq) \vec{ S}_{\G l}(\bk ,\bk+\bq)$
is the orbital--weighted spin operator for the pocket $l=X,Y$, where 
$ \vec{S}_{\G l}(\bk,\bk+\bq)=\frac{1}{2}\sum_{\a \b}  d^\dagger_{\G \bk \a} \vec{\s}_{\a \b }d_{l \bk + \bq \b}$. 
The weight factors $w^\eta_{\G l}(\bk,\bk + \bq)=a_{\G \eta}(\bk) a_{l \eta}(\bk+\bq)$, 
relate orbital and pockets basis, while $\vec{\sigma}_{\alpha\beta}$ are the Pauli matrices. 

Let us now introduce the bosonic fields $\vec{\Delta}_{\eta l=X,Y}$ associated 
to the magnetic d.o.f  $\vec{S}_{\eta l=X,Y}$. Via a standard 
Hubbard--Stratonovich (HS) machinery we derive the effective action up to the 
quartic order (for a complete derivation see Appendix \ref{sect:appA}):
\begin{eqnarray}
S_{\text{eff}}&=&\frac{1}{2}\sum_{l=X,Y}\sum_{\eta_1\eta_2}r_{\eta_1,\eta_2,l}\vec{\Delta}_{\eta_{1}l}
\cdot\vec{\Delta}_{\eta_{2}l}+\nonumber\\
&+&\frac{1}{4}\sum_{\eta_{1}\eta_{2}\eta_{3}\eta_{4}} \frac{1}{2} u_{\eta_{1}\eta_{2}\eta_{3}\eta_{4}}
\psi_{\eta_{1}\eta_{2}}\psi_{\eta_{3}\eta_{4}}+\nonumber \\
&-&\frac{1}{2}g_{\eta_{1}\eta_{2}\eta_{3}\eta_{4}}\phi_{\eta_{1}\eta_{2}}\phi_{\eta_{3}\eta_{4}}
+v_{\eta_{1}\eta_{2}\eta_{3}\eta_{4}}\psi_{\eta_{1}\eta_{2}}\phi_{\eta_{3}\eta_{4}} 
\label{eq:Seff}
\end{eqnarray}
where $\vec{\Delta}_{\eta_{1}X}$, $\vec{\Delta}_{\eta_{1}Y}$ are the OPs with ordering 
momentum either ${\bf Q}_1=(\pi,0)$ or ${\bf Q}_2=(0,\pi)$ in the Landau action $S_{\text{eff}}$, 
and we decomposed the quartic term using the following parametrization
\begin{equation}
\psi_{\eta_{1}\eta_{2}}=2\left(\vec{\Delta}_{\eta_{1}X}\cdot\vec{\Delta}_{\eta_{2}X}+
\vec{\Delta}_{\eta_{1}Y}\cdot\vec{\Delta}_{\eta_{2}Y}\right),
\label{eq:psi}
\end{equation}
\begin{equation}
\phi_{\eta_{1}\eta_{2}}=2\left(\vec{\Delta}_{\eta_{1}X}
\cdot\vec{\Delta}_{\eta_{2}X}-\vec{\Delta}_{\eta_{1}Y}
\cdot\vec{\Delta}_{\eta_{2}Y}\right),
\label{eq:orderparameter}
\end{equation}
where $\phi_{\eta_1 \eta_2}$ is the nematic field in our approach. 
For a more general parametrization see Appendix \ref{sect:appA}.

The Landau coefficients are given by
\begin{subequations}
\begin{equation}
r_{l\eta_1\eta_2}={U^{\text{spin}}_{\eta_1\eta_2}}^{-1}+\frac{1}{4}\sum_{k}G_\Gamma G_l 
w_{\Gamma l}^{\eta_1} w_{\Gamma l}^{\eta_2}, \label{eq:quadratic}
\end{equation}
\begin{equation}
u_{\eta_{1}\eta_{2}\eta_{3}\eta_{4}}=\frac{1}{2}\sum_{kl}G^{2}_{\Gamma}
\left(G_{l} w^{\eta_{1}}_{\Gamma l} w^{\eta_{2}}_{\Gamma l}\right)
\left(G_{l} w^{\eta_{3}}_{\Gamma l} w^{\eta_{4}}_{\Gamma l}\right),
\label{eq:u}
\end{equation}
\begin{equation}
g_{\eta_{1}\eta_{2}\eta_{3}\eta_{4}}=-\frac{1}{2}\sum_{kl} \sum_{s=1(X),-1(Y)}G^{2}_{\Gamma}
\left(s\, G_{l} w^{\eta_{1}}_{\Gamma l} w^{\eta_{2}}_{\Gamma l}\right)
\left(s\, G_{l} w^{\eta_{3}}_{\Gamma l} w^{\eta_{4}}_{\Gamma l}\right),
\label{eq:g}
\end{equation}
\begin{equation}
v_{\eta_{1}\eta_{2}\eta_{3}\eta_{4}}=\frac{1}{2}\sum_{kl} \sum_{s=1(X),-1(Y)}G^{2}_{\Gamma}
G_{l} w^{\eta_{1}}_{\Gamma l} w^{\eta_{2}}_{\Gamma l}
\left(s \, G_{l} w^{\eta_{3}}_{\Gamma l}w^{\eta_{4}}_{\Gamma l}\right),
\label{eq:v}
\end{equation}
\end{subequations}
with $l=X,Y$. $G_{m, k} = (i \o_n - \xi_{m,\bk})^{-1}$, $m=\G, X, Y$ are the non 
interacting single--particle Green's functions. The spin OP $\vec{\Delta}_{\eta l=X,Y}$ 
is a vector in the orbital space and the effective coefficients of Eq.~\pref{eq:Seff} 
are matrix/tensor in the same space. If we consider the weight factors 
$w^{\eta}_{\Gamma l}=1$ we recover the effective action given by Eq.~(7) 
and the Landau coefficients given by Eq.~(8) obtained via the orbital--less 
approach of [\onlinecite{schmalianprb12}]. 

The effective action, Eq.~\pref{eq:Seff}, is invariant under the O(3) symmetry 
and under the interchange between $\vec{\Delta}_X$ and $\vec{\Delta}_Y$. The 
quadratic coefficient, Eq.~\pref{eq:quadratic}, is the $\eta_1 \eta_2$ component 
of the quadratic Landau parameter $\hat{r}_{l=X,Y}$. The magnetic susceptibility 
is defined by $\hat{r}_{l} (\bq, i\O_m)= \hat{\chi}^{-1}_{l}(\bq, i\O_m)$ and 
the N\'eel temperature $T_N$ is fixed by the divergence of $\hat{\chi}_l(0,0)$. 
Due to the presence of the orbital d.o.f. the $T_N$ obtained within this 
formulation is a function of $(U, J_H)$. The quartic Landau coefficients 
Eq.~\pref{eq:u}--\pref{eq:v} are the elements of the $\hat{u}$, $\hat{g}$ and 
$\hat{v}$ tensors. The $\hat{g}$ tensor coupled to the 
$(\Delta_X^2-\Delta_Y^2)$ term is the nematic coupling. By minimizing the action 
Eq.~\pref{eq:Seff} two solutions for $T<T_N$ corresponding to the $(\pi,0)$ and 
$(0,\pi)$ magnetic state can be found in a proper range of parameters. To 
analyze the nematic phase we need to study how the $\hat{\phi}$ field orders. In 
[\onlinecite{schmalianprb12}], the nematic OP is related with the magnetic 
susceptibility via a second HS transformation in terms of $\hat{\psi}$ and 
$\hat{\phi}$ given in Eq.~\pref{eq:psi} and Eq.~\pref{eq:orderparameter}. A 
detailed study of the nematic transition goes beyond the aim of this work and it 
will be the subject of further investigations \cite{workinprogress}. We focus on 
the analysis of the Landau parameters of the effective action Eq.~\pref{eq:Seff} 
that allows us to demonstrate that taking into account the orbital d.o.f changes 
the S--N picture drastically. 

First let us focus on the quartic coefficients controlling the nematic order. 
While within the orbital--less Landau approach \cite{schmalianprb12} a finite $g$ 
requires the e\,--\,pockets to be elliptical \cite{chubukovprb10, schmalianprb12, 
Fernandesreview12} and $v$ is zero unless interaction between e\,--\,pockets are 
taking into account \cite{nematicnatcomm14}, this is no longer true in the 
present formulation. In fact, regardless of ellipticity, both $\hat{g}$ and 
$\hat{v}$, Eqs.\pref{eq:g}--\pref{eq:v}, are finite due to the $yz/zx$ orbital 
content of the X/Y e\,--\,pockets i.e. $w^{yz}_{\Gamma Y} \sim 0$,  
$w^{zx}_{\Gamma X} \sim 0$. This result holds for any number of orbitals 
since it follows from symmetry arguments. This general outcome implies that the 
nematicity does not necessarily require the ellipticity of the e\,--\,pockets as 
expected in the standard S--N scenario. Below we will address the question about the relative 
contribution of the orbital C4 symmetry and the ellipticity 
of the e\,--\,pockets to the nematic coupling analyzing the simple case of 
the continuum limit of a two-orbital model.

\begin{figure}[t]
\centering
\includegraphics[clip,width=0.48\textwidth]{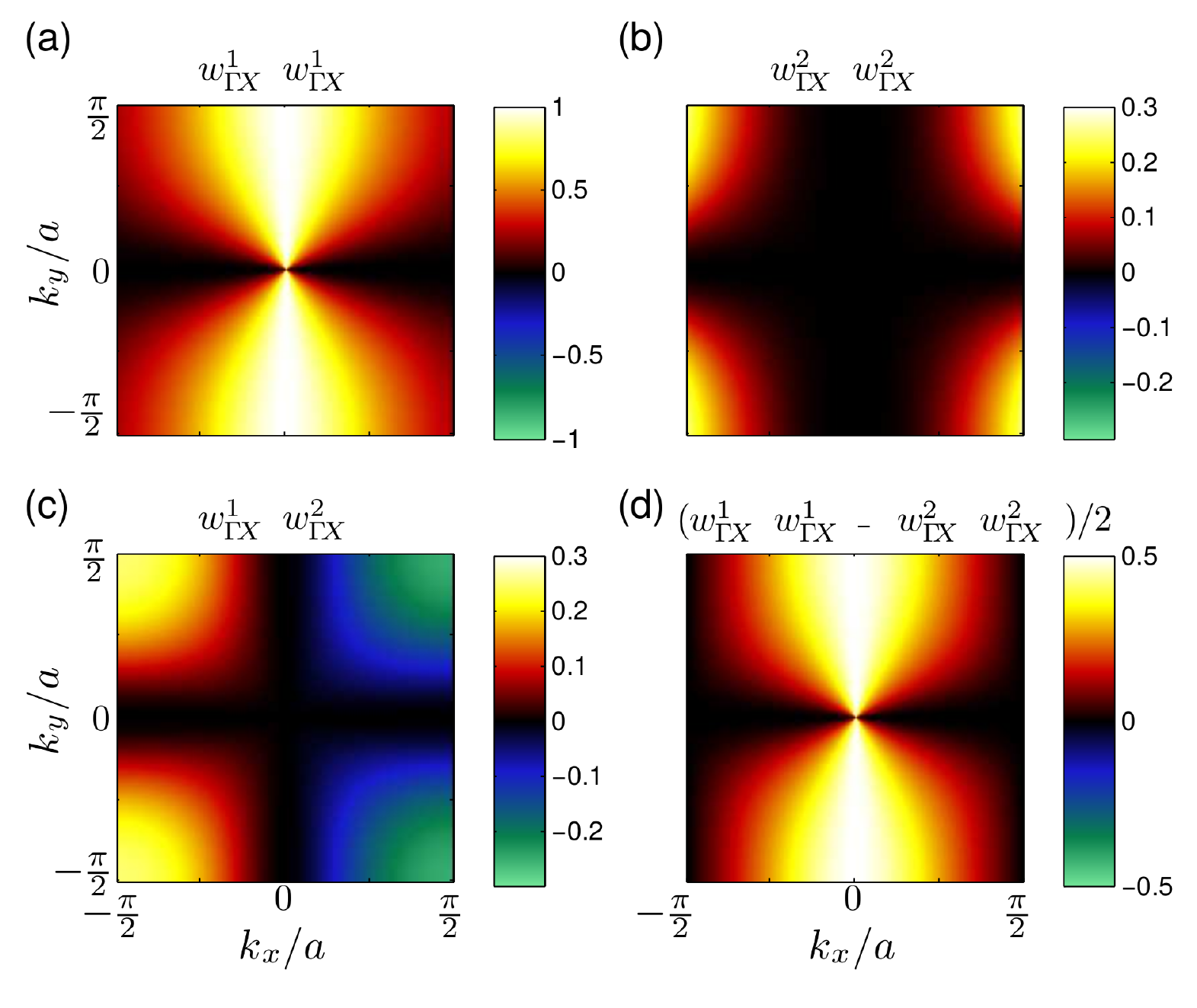} 
\caption{(Color online) (a--c) $w_{\Gamma X}^{\eta_1}w_{\Gamma 
X}^{\eta_2}$ factors weighting the {\bf k}--integrals of the magnetic bubble 
Eq.~\pref{eq:bubble} computed using the same model parameters of [\onlinecite{graser09}]. 
The dominant orbital component is the $w_{\G X}^{1} 
w_{\G X}^{1}$. The contribution from the orbital 2 is considerably smaller as 
well as the mixed one $w_{\G X}^{1} w_{\G X}^{2}$.
By drawing the $\bk$--dependence of the orbital weights for the magnetic 
fluctuations $\Gamma Y$ we would find the same results rotated by $\pi/2$ as 
expected by symmetry. (d) $\tau_3$ weight component $w_{\Gamma 
X}^3=\frac{1}{2}(w_{\Gamma X}^{1}w_{\Gamma X}^{1}-w_{\Gamma 
X}^{2}w_{\Gamma X}^{2})$. The $\tau_3$ weight is similar to $\sim 
w_{\Gamma X}^{1}w_{\Gamma X}^{1}/2$ as found in the simplified model 
where we assume the contribution of the orbital $2$ being zero for the 
$X$--pocket.} 
\label{fig:weights}
\vspace{-0.3cm}
\end{figure}

Next, we analyze the $(\pi,0)$ magnetic susceptibility in the nematic phase
defined by Eq.~\pref{eq:quadratic} as
\begin{eqnarray}
\label{eq:complete_susceptibility}
\chi^{-1}_{X \, \eta_1\eta_2}({\bf q})&=& {U^{\text{spin}}_{\eta_1\eta_2}}^{-1} +  \frac{1}{2}\Pi_{X}^{\eta_1\eta_2}(q),  \\ 
\Pi_{X}^{\eta_1\eta_2}(q) &=& \frac{1}{2}\sum_{k} G_{\Gamma \, k}  G_{X \, k+q}  \nn \\ &&
w_{\Gamma X}^{\eta_1}({\bf k, k+q})w_{\Gamma X}^{\eta_2}({\bf k,k+q}),
\label{eq:bubble}
\end{eqnarray}
where $k\equiv (i\o_n, \bk)$,  $q \equiv(i\O_m, \bq)$, with $\o_n$, $\O_m$ 
fermion, boson Matsubara frequencies respectively. We restrict our analysis to a 
two--orbital model\cite{graser09, nosotrasnjp09,dunghailee09}
$\eta_{1,2} = yz,zx$. For simplicity we will use $yz=1$ and $zx=2$. The weight 
factor $w_{\Gamma X}^{\eta_1}w_{\Gamma X}^{\eta_2}$ numerically 
computed for the $d_{yz}$--\,$d_{zx}$ model are shown in 
Fig.~\ref{fig:weights}(a--c). 
The strong $\bk$--dependence of the orbital weights modulates the $\bk$--integral 
in Eq.~\pref{eq:bubble}. Notice that the contributions to the susceptibility 
along the $k_x$ direction have zero weight for all orbital components. 
$\Pi_X^{11}$ is the dominant orbital component of the magnetic bubble with the 
others being suppressed by the small orbital weights.

Let us consider the constant part $q=0$ of the magnetic susceptibility. By using 
Pauli matrices basis to decompose the orbital structure of 
Eq.~\pref{eq:complete_susceptibility}, we find $\hat{\chi}^{-1}_{X} (0) \equiv  
\big( \tilde{U} + \Pi^0_X \big) \hat{\tau}_{0} -  \big(\tilde{J}_H - \Pi^1_X 
\big) \hat{\tau}_1 +  \Pi^3_X \hat{\tau}_{3}$, where $\Pi^{0,3}_X=(\Pi_{X}^{11} \pm \Pi_{X}^{22})/4$ 
and $\Pi^1_X=\Pi_{X}^{1 2}/2$. $\tilde{U} = (8U/3)/\text{det}U^{\text{spin}}$, 
$\tilde{J}_H= 4J_H/\text{det}U^{\text{spin}}$, where $\text{det}$ indicates the 
determinant.
The $\tau_0,\, \tau_1$ components of the magnetic bubble  
renormalize $U$ and $J_H$ coupling to a smaller value. Interestingly, a 
non--zero $\tau_3$ component reflecting the breaking of the orbital degeneracy 
is spontaneously generated, see Fig.~\ref{fig:weights}(d).
The existence of a finite $\tau_3$ component in the spin sector is connected to 
the orbital splitting in the charge sector. This can be demonstrated deriving 
the effective action $S_{\text{eff}}\,[n_{\eta m},\Delta_{\eta l}]$ that accounts 
also for the charge d.o.f. at small momentum, $n_{\eta m}$, with $m=\Gamma, X, 
Y$, and  $l=X, Y$. In the nematic phase, the charge susceptibility computed at 
Gaussian level, $n_{\eta \Gamma}^2$, does not present $yz$/$zx$ orbital 
splitting. However, at higher level of the effective action, there are terms 
coupling the spin mode with broken orbital degeneracy to the charge mode (e.g. 
$(n_{\eta \Gamma} \Delta_{\eta X})^2$). Integrating out the spin fluctuations, a 
$\tau_3$ component is induced in the charge field susceptibility i.e. orbital 
splitting is generated into the charge sector. This intrinsic splitting arises 
because of the effect of the non trivial $yz/zx$ $C_4$ symmetry at low energy. 
This mechanism is different from breaking explicitly the $yz$/$zx$ symmetry by 
introducing a small crystal field in the Hamiltonian \cite{schmalianprb12, 
lvphillipsprb10} or from inferring it assuming X--Y pocket interaction 
\cite{stephanhaasprb13}.  Our analysis clearly reveals the intrinsic 
interrelation between spin and orbital d.o.f. in the nematic phase and it 
extends the MF findings of orbital splitting in the magnetic phase 
\cite{leeyinku09, yu09, nosotrasprl10, nosotrasprl10-2, daghofer10, 
yinpickettprb10, yin11}. 

Finally we analyze the momentum dependence of the spin susceptibility 
$\hat{\chi}(\bf q)$, Eq.~\pref{eq:complete_susceptibility}. For perfectly nested 
pockets and without the orbital structure one obtains isotropic spin 
fluctuations $\Pi_{X}(0,\mathbf{q})=\Pi_{X}(0)+c_X^x  q_x^2+c_X^y  q_y^2$ where 
$c_X^i\sim (\partial_{k_x} \xi_\bk)^2$. In the present case the expansion also 
involves the $\bk$--dependent orbital weight factors 
(Fig.~\ref{fig:weights}(a--c)). Due to the singular orbital symmetry of the 
problem we find that the angular dependence introduced by the weights in the 
magnetic bubble leads to anisotropic component in the spin fluctuations $c_X^x 
\neq c_X^y$. We discuss quantitatively these results below within a simplified model 
obtained by the continuum limit of the two-orbital model. 

\section{QUANTITATIVE DISCUSSION}
Our orbital--dependent analysis of the Landau parameters changes qualitatively 
the S--N picture derived in the orbital--less approach. We have shown that 
ellipticity is mandatory neither to find a finite nematic coupling $\hat{g}$ nor 
to have anisotropic spin fluctuations. However, since the e\,--\,pockets are 
elliptical we quantify the relative importance of these effects.

We take the continuum limit of the two--orbital model that provides us with 
simple analytical expressions to be compared to the orbital--less effective 
action. For simplicity we 
assume that the X(Y) pocket are exclusively composed by the $1$($2$) orbital.  
Then all the weight factors vanish except $w_{\G X}^{1} w_{\G X}^{1} = 
\sin^2{\theta_\bk}$ and $w_{\G Y}^{2}w_{\G Y}^{2}=\cos^2{\theta_\bk}$ with 
$\theta_\bk=\arctan(k_y/k_x)$. These terms reproduce the ${\bk}$--dependence 
weights numerically computed in Fig.~\ref{fig:weights}(a--c). The parameter 
$\delta m$, that encodes the ellipticity, is introduced perturbatively via $ 
1/m_{x/y} = (1 \pm \delta m)/m$ with $m_{x/y}$ the e--pocket mass. 
More details about the model can be found in Appendix \ref{sect:appB}.

First we analyze the momentum dependence of the magnetic fluctuations. Within 
this approximation the magnetic bubble $\hat{\Pi}_X(q)$ Eq.~\pref{eq:bubble} has 
only one finite orbital component 
\begin{eqnarray}
\Pi_{X}^{11}(q) = \frac{1}{2}\sum_n\int \frac{d^2 \bk}{(2\pi)^2}  
\frac{ \sin^2\theta_{\bk} }
{(i\omega_n-\xi_{X, \bk})(i\o_n-\xi_{\G,\bk+\bq})}
\label{eq:lareina}
\end{eqnarray}
By expanding up to the leading order in momentum we find anisotropic stiffness 
due to the angular dependence introduced by the $w^1_{\Gamma X}w^1_{\Gamma X} 
\sim \sin^2\theta_{\bk}$: $c_X^x-c_X^y \sim (1-4 \delta m)v_F^2N_F/T^2$ with 
$v_F$ the Fermi velocity and $N_F$ the density of states at the Fermi level (see 
Appendix \ref{sect:appB} for furthers details). The $\sin^2\theta_{\bk}$ factor 
comes from the topological feature of the spin interaction connecting the 
$\Gamma$ pocket with vorticity 2 with the  topologically trivial X pocket. In 
the orbital--less effective action $c_X^x-c_X^y  \sim (-8\delta m)v_F^2N_F/T^2$. 
Therefore the topological feature affects the anisotropic stiffness already at 
zero order and in addition reduces the contribution proportional to $\d m$ that 
is opposite in sign.

We turn now to the calculation of the Landau parameters. Focusing on the quartic 
terms of the effective action we notice that, within this approximation, only 
few components of the $\hat{g}$, $\hat{u}$, $\hat{v}$ tensor 
are finite. Moreover due to the angular symmetry it holds: 
$u_{2222}=u_{1111}= - g_{1111}$, 
$u_{1122}=u_{2211} = g_{1122}$, $v_{1111}= - v_{2222}= - g_{1111}$, 
$v_{1122}= v_{2211}= - g_{1122}$ and $g_{1111}=g_{2222}$, $g_{1122}=g_{2211}$ 
leaving just two independent components.
At finite temperature we find 
\begin{equation}
g_{11\eta\eta}=\frac{N_F}{\pi T^2}\;a_\eta \bigg( 1+b_\eta\;\frac{\epsilon_0^2}{T^2} \delta m^{2}\bigg)\; ; 
\label{eq:final_g}
\end{equation}
with $a_\eta$ and $b_\eta$ ($\eta=1,2$) numerical factors (see Appendix \ref{sect:appB}). 
The nematic coupling is finite already at zero order due to the orbital symmetry. 
The first finite term proportional to $\delta m$ appears at the second order and 
it is suppressed by temperature with respect to the zero--order contribution. If 
the ellipticity is taken into account perturbatively, as usually in the 
literature, then the effect of the orbital symmetry turns out to be the dominant 
one. On the other hand, if we are not allowed to expand in $\d m$, the result of 
Eq.~(\ref{eq:final_g}) would change and both the effects would contribute 
on equal footing. 
Notice however that the orbital symmetry is an intrinsic property 
of the system always present whereas the ellipticity depends on details of the 
FSs.

\section{CONCLUSION}
We have derived the effective action for the spin order parameter with Landau 
coefficients depending on the orbital character, Hubbard and Hund's 
interaction. 
This derivation can be adapted for any multiorbital system with spin--orbital 
interplay. In the context of iron superconductors, this model allows for a good 
description of the non--trivial low energy spectra of the system. Important 
differences are revealed with respect to the orbital--less effective approach: 
(i) the nematic coupling is finite due to the $yz/zx$ orbital content of the $X$ and 
$Y$ pockets regardless of the value of the ellipticity of the e--pocket. 
(ii) In the nematic phase the $yz$/$zx$ degeneracy is spontaneously broken in the 
spin channel due to the non--trivial $C_4$ symmetry. This broken symmetry induces 
orbital splitting in the charge channel revealing the strong spin--orbital 
interplay. (iii) The spin fluctuations have anisotropic momentum dependence in 
the $x$/$y$ directions even in the case of perfect nesting due to the presence 
of the vortex at the $\Gamma$ pocket. The vortex also affects the ellipticity 
contribution. In the light of our results, experiments interpreted in terms of 
the ellipticity of the e\,--\,pockets should be revisited including the orbital 
degree of freedom. We have quantified the general results via the analytical treatment of the 
continuum limit of the two orbital model. Since our conclusions are based in 
topological and symmetry arguments, they are also valid for more realistic 
models describing iron pnictides. This work is a necessary bridge between 
multiorbital physics and effective theories of spin fluctuations.

We thank M.A.H. Vozmediano, E. Fradkin and A. Chubukov for useful discussions 
and L. Benfatto, E. Bascones and M.J. Calder\'on for a critical reading of the 
manuscript. B.V. acknowledges the Fundaci\'on Ram\'on Areces and the Ministerio de 
Econom\'ia y Competitividad through Grant No. FIS2011-29689 and Grant No. 
FIS2014-53219-P for financial support. L.F. acknowledges funding from the 
University of Rome ``Sapienza''. A. C. acknowledges the JAE-doc program and the 
Ministerio de Ciencia y Educacion through Grants No. FIS2011-23713 and No. 
PIB2010BZ-00512 for financial support.

\appendix
\section{MICROSCOPIC CALCULATION OF THE EFFECTIVE ACTION} 
\label{sect:appA}

The straightforward way to obtain the action in terms of the spin excitations 
$\vec{\Delta}_{\eta  X}$, $\vec{\Delta}_{\eta  Y}$ is by means of the 
Hubbard-Stratonovich (HS) transformation. Here we summarize the main steps 
of the derivation of the effective action $S_{\text{eff}}$ in Eq.~\pref{eq:Seff}. 
More details about the HS procedure can be found in Ref.s 
[\onlinecite{negele}] and [\onlinecite{benfatto04}] and references therein.

Starting from the microscopic 
Hamiltonian we can define the microscopic action $S[c_i(\t)]$ as
\be
\lb{eq:actmicro}
S[c_i(\t)] = \int_{0}^{\beta}{d\t c^{\dagger}_i(\t)[\partial_{\t} - \m]c_i(\t) +
\hat{H}[c_i(\t)]},
\ee
$\t$ is the imaginary time and $\beta=1/T$. 
Then the partition function of our system can be computed as the integral over
Grassmann variables, 
\be
\lb{eq:partition}
Z = \int \DD c  \, e^{-S[c_i(\t)]}.
\ee
The HS transformation allows us to decouple the quartic term in fermionic operators
via the functional identity
$$
e^{\frac{\pm a x^2}{2}} = \int  \DD y  e^{ -\frac{y^2}{2a} + \sqrt{\pm} y x} \quad a>0
$$
with $y$ an auxiliary variable, i.e. the HS field. 
At this point the action becomes quadratic with respect to the fermionic 
operators so that  we can integrate out the fermions from 
Eq.~\pref{eq:partition}. The results of this operation is recast back into the 
exponent and the partition function is expressed  in terms of the effective 
action $S[y]$ 
\be
\lb{Z}
Z = \int  \DD y e^{-S[y]}.
\ee  

We need to specialize this machinery to the analysis of our case. 
Pnictide systems have to be described using a multiorbital Hamiltonian.
Typical hamiltonians include the tigh-binding $t^{\eta_1\eta_2}_{ij}$, 
the crystal field splitting $\epsilon_{\eta_1}$
and local interactions restricted to the Fe orbitals: 
the in\-tra-or\-bi\-tal Hubbard's coupling $U$, the interorbital one $U^\prime$, 
the Hund's coupling $J_H$ and the pairing term $J^\prime$:
\begin{eqnarray}
& H &  = 
\sum_{ij\eta_1\eta_2\alpha}t^{\eta_1\eta_2}_{ij}c^\dagger_{i\eta_1\alpha}c_{j\eta_2\alpha}+h.c.
+ U\sum_{j\eta_1}n_{j\eta_1\uparrow}n_{j\eta_1\downarrow}  \nn \\ 
& +&  (U'-\frac{J_H}{2})\sum_{j\eta_1>\eta_2;\alpha\beta}n_{j\eta_1\alpha}n_{j\eta_2\beta}
-2J_H\sum_{j\eta_1 >\eta_2}\vec{ S}_{j\eta_1}\cdot \vec{ S}_{j\eta_2} \nn \\
& + &  J'\sum_{j\eta_1\neq\eta_2}c^\dagger_{j\eta_1\uparrow}c^\dagger_{j\eta_1\downarrow}
c_{j\eta_2\downarrow}c_{j\eta_2\uparrow}
- \sum_{j\eta_1\alpha} ( \mu -\epsilon_{\eta_1}) n_{j\eta_1 \alpha} \,
\label{eq:hamiltoniano}
\end{eqnarray}
$i,j$ label the Fe sites in the Fe unit cell, $\alpha$ and $\beta$ the spin and 
$\eta_1$, $\eta_2$ the Fe d-orbitals. $ c^{(\dagger)}_{i \eta_1 \alpha}$ 
destroys (creates) a fermion  with spin $\alpha$ in the $\eta_1$ orbital on the 
$i$-th site, $n_{i \eta_1 \alpha} = c^\dagger_{i\eta_1 \alpha}c_{i\eta_1 
\alpha}$ and $\vec{ S}_{i \eta_1} = \frac{1}{2}\sum_{\a \b}  c^\dagger_{i \eta_1 
\a} \vec{\s}_{\a \b } c_{i \eta_1 \b}$, with $\vec{\sigma}_{\alpha\beta}$ the 
Pauli matrices. Assuming our system rotationally invariant we can use 
$U'=U-2J_H$ and $J'=J_H$ leaving only two independent parameters, $U$ and $J_H$ 
~\cite{castellani78}. Repulsion between electrons requires $J_H<U/3$. 

Transforming by Fourier and changing the basis from the orbitals to the 
bands, the tigh-binding term can be diagonalized. We are interested in the low 
energy physics thus we restrict to energies and momenta close to the FS taking 
into account the h\,--\,pocket at $\Gamma$ and the $X$ and $Y$ e\,--\,pockets. 
We consider only the spin channel of the interaction of the Hamiltonian and we 
restrict to the spin excitations with momenta near ${\bf Q}_1=(\pi,0)$ and ${\bf 
Q}_2=(0,\pi)$. Since the pair-hopping term $J'$ does not contribute to the spin 
channel we will not have it into account. 
The complete microscopic Hamiltonian reads 
\bea
H  &=&\sum_{m=\G, X,Y} \sum_{\bk \a} \xi_{m \bk } d_{m \bk \a }^\dagger d_{m \bk \a } = \nn \\ 
&& - \frac{1}{2}\sum_{l=X,Y} \sum_{\bf q}\sum_{\eta_1\eta_2} 
U^{\text{spin}}_{\eta_1\eta_2}\vec{ S}_{\eta_1l}({\bf q})\cdot\vec{ S}_{\eta_2l}({-\bf q}).
\eea
Here $ d^{(\dagger)}_{m \bk \a }$ destroys (creates) a fermion with momentum 
$\bk$ and spin $\alpha$ in the $m$ pocket, $\xi_m=\e_m-\mu$ where $\e_m$ is the 
fermionic dispersion. The spin coupling is
$U^{\text{spin}}_{\eta_1\eta_2}=\frac{8}{3}U\delta_{\eta_1\eta_2} + 
4J_H(1-\delta_{\eta_1\eta_2})$, the orbital--weighted 
spin operator for the pocket $l=X,Y$ is 
$$\vec{ S}_{\eta l}(\bq)=\sum_\bk w_{\G l}^\eta(\bk,\bk + \bq) \vec{ S}_{\G l}(\bk ,\bk+\bq)$$
where 
$$ \vec{S}_{\G l}(\bk,\bk+\bq)=\frac{1}{2}\sum_{\a \b}  d^\dagger_{\G \bk \a} \vec{\s}_{\a \b 
}d_{l \bk + \bq \b}.$$ 
The weight factors $w^\eta_{\G l}(\bk,\bk + \bq)=a_{\G \eta}(\bk) a_{l \eta}(\bk+\bq)$, 
relate orbital and pockets basis. 

We introduce $\Psi^\dagger, \Psi$  six-dimensional creation, destruction operators
\be
\lb{eq:Nambu}
\Psi^\dagger_\bk = 
( d^\dagger_{\G, \bk\uparrow} \  d^\dagger_{\G, \bk\downarrow}  \
d^\dagger_{X, \bk\uparrow} \  d^\dagger_{X, \bk\downarrow}  \   
d^\dagger_{Y, \bk\uparrow} \  d^\dagger_{Y, \bk\downarrow}  ). 
\ee 

The auxiliary bosonic fields $\vec{\Delta}_{\eta l}$ coupled to $\vec{ S}_{\eta 
l}({\bf q})$ are our HS fields and play the role of the magnetic order 
parameters in the Landau functional. Although the presence of a finite 
interorbital coupling, the $U^{\text{spin}}$ matrix is positive defined within 
the range of validity of the model, $J_H < U/3$, thus we can apply the standard 
HS transformation. If $U^{\text{spin}}$ had develop negative eigenvalues the HS 
decoupling would lead to the appearance of an imaginary unit in the effective 
action that have to be handled properly. For further details we refer to 
[\onlinecite{laura09}] and [\onlinecite{laura13}] where this problem has been a\-na\-ly\-zed in the 
context of the superconducting instability. 

After the HS transformation the partition function can be written as
\be
\lb{eq:Zcomplete}
Z = \int \DD \Psi \DD \D e^{-S[\Psi,  \D]},
\ee  
with
\bea
\lb{eq:Scomplete}
S[\Psi,  \D] &= & 
\frac{1}{2} \sum_{q} {U^{\text{spin}}_{\eta_1 \eta_2}}^{-1} \sum_{l=X,Y} \vec{\D}_{\eta_1 l, q }\vec{\D}_{\eta_2 l, - q }  \nn \\ 
&& \nn \\
&& +  \sum_{k k^{\prime}} \Psi^{\dagger}_k \ \hat{A}_{k k^\prime} \ \Psi_{k^\prime} \ .
\eea
We used $k-k^\prime = q$, $k\equiv(\bk,i\o_n)$, $q\equiv(\bq,i\O_m)$, with 
$\o_n$, $\Omega_m$ Matsubara fermion and boson frequencies, respectively. 
Hereafter the repeated orbital indices are 
summed. The $\hat{A}_{k k^\prime}$ matrix is composed by the blocks:
\bea
\lb{eq:Amm}
&& \hat{A}_{k,k^\prime}|_{m m} = -\hat{G}^{-1}_{m, k k^\prime}\delta_{k k^\prime}  \\
&& \hat{A}_{k,k^\prime}|_{\G \, l} = \ \frac{1}{4}\hat{\vec{\D}}_{\eta l, k - k^\prime} \, w^{\eta}_{\G l} (\bk, \bk^\prime)
\lb{eq:Agl}
\eea
%
%
where we use a compact notation for the spin sector defining  
$$\hat{G}_{m, k} = G_{m, k}\cdot {\bf  I}, \quad \quad \quad 
\hat{\vec{\D}}_{\eta l, q} = \vec{\D}_{\eta l, q} \cdot \vec{{\boldsymbol \sigma}}$$ 
with ${\bf  I}$ the identity matrix ${\bf I}$ 
and  $ {\boldsymbol \sigma}^i$ the Pauli matrices. $G_{m, k} = (i \o_n - 
\xi_{m \bk})^{-1}$ are the non interacting single-particle Green's functions. 

The exact integration of the fermionic d.o.f in Eq.~\pref{eq:Zcomplete} 
gives us an expression for the effective action in terms of the HS field
$\D_{\eta l }$ only
\be
\label{eq:SDonly}
S_{eff} =   \frac{1}{2} \sum_{q} {U^{\text{spin}}_{\eta_1 \eta_2}}^{-1} \sum_{l=X,Y}
 \vec{\D}_{\eta_1 l, q }\vec{\D}_{\eta_2 l, - q } - \mathrm {Tr}\log {\hat{A}_{kk^\prime}},
\ee
It is convenient decompose the second term of Eq.~\pref{eq:SDonly} by 
separating in Eq.s~(\ref{eq:Amm}-\ref{eq:Agl}) the part with the explicit structure of $\delta_{k k'}$ 
from the rest $\hat{A}_{kk^\prime} = -\hat{{\cal G}}_0^{-1} + \hat{\VV}_{kk^\prime} $. 
This separation allow us to rewrite in our action
\bea
\mbox{Tr}\ln{A_{kk'}} =  \mbox{Tr}\ln{\hat{{\cal G}}_0^{-1}}+ 
\mbox{Tr}\ln{\left[\hat{1}-\hat{{\cal G}}_{0}\hat{\VV}\right]}
\label{eq:Tr}
\eea

Now we separate the HS fields 
$\vec{\D}_{\eta l, q}=\vec{\D}_{\eta l, 0}+ \d \vec{\D}_{\eta l, q}$ 
in its homogeneous and constant part and its fluctuating part. 
By minimizing the action with respect to $\vec{\D}_{\eta l, 0}$ 
we obtain the mean-field (MF) equations of the magnetic problem 
that admit non trivial solutions with finite magnetization below 
a critical temperature. 

Beyond MF we need to take into account the fluctuation of the magnetic 
HS fields around their MF value. It is easy to verify that we 
can expand 
$$
\mbox{Tr}\ln{\left[\hat{1}-\hat{{\cal G}}_{0}\hat{\VV}\right]} = 
\sum_{n}\frac{1}{n}\mathrm{Tr}[ \hat{\GG}_0 \hat{\VV}_{k-k'}]^n,
$$
in Eq.~\pref{eq:Tr}.
We are interested into the nematic d.o.f. so that we need to retain terms up to 
the quartic order of the expansion in the magnetic HS fields. 
After a bit of algebra we obtain the effective action in the non-magnetic phase:
\bea
S_{eff} &=& \frac{1}{2}\sum_{l=X,Y}  \bigg( {U^{spin}_{\eta_1 \eta_2}}^{-1}  +  
\frac{1}{2} \Pi_{l}^{\eta_1 \eta_2} \bigg) \vec{\D}_{\eta_1 l} \cdot \vec{\Delta}_{\eta_2 l} + \nn \\ 
&&+ \frac{1}{16} \sum_{l, l^\prime = X,Y} \l_{l l^\prime}^{ \eta_1 \eta_2 \eta_3 \eta_4}  \left( \vec{\D}_{\eta_1 l} \cdot \vec{\D}_{\eta_2 l} \right)
\left(\vec{\D}_{\eta_3 l^\prime} \cdot \vec{\D}_{\eta_4 l^\prime} \right) \nonumber \\
\label{eq:Sfinal}
\eea 
where $\Pi_{l}^{\eta_1\eta_2} $ is defined as
\be
\Pi_{l}^{\eta_1\eta_2} = \frac{1}{2}\sum_{k} G_{\Gamma}  G_{l}   
w_{\Gamma l}^{\eta_1} w_{\Gamma l}^{\eta_2}, 
\ee
the $\l_{l l^\prime }^{\eta_1 \eta_2 \eta_3 \eta_4}$  are given by
\bea
\lambda_{X X}^{ \eta_1 \eta_2 \eta_3 \eta_4}  &=& \frac{1}{16}
\sum_{iw,\mathbf{k}}G^{2}_{\Gamma}G^{2}_{X}
w^{\eta_{1}}_{\G X}w^{\eta_{2}}_{\G X}w^{\eta_{3}}_{\G X}w^{\eta_{4}}_{\G X},  \nn \\
\lambda_{Y Y}^{ \eta_1 \eta_2 \eta_3 \eta_4}  &=& \frac{1}{16}
\sum_{iw,\mathbf{k}}G^{2}_{\G}G^{2}_{Y}
w^{\eta_{1}}_{\G Y}w^{\eta_{2}}_{\G Y}w^{\eta_{3}}_{\G Y}w^{\eta_{4}}_{\G Y},  \nn \\
\lambda_{X Y}^{ \eta_1 \eta_2 \eta_3 \eta_4}  &=& \frac{1}{8}
\sum_{iw,\mathbf{k}}G^{2}_{\G}G_{X}G_{Y}
w^{\eta_{1}}_{\G X}w^{\eta_{2}}_{\G X}w^{\eta_{3}}_{\G Y}w^{\eta_{4}}_{\G Y}. \nonumber \\
\eea
and we simplified a bit the notation using $\d \vec{\D}_{\eta l} \ra  \vec{\D}_{\eta l}$ and 
dropping the $k, q$ dependencies of the variables.
At this point  it is already recognizable the definition for the $r$ coefficient 
of the gaussian part. For the quartic part we need to define properly the $\psi, \phi$ operators. 
The more general parametrization would be 
\bea 
\label{eq:gen_psi}
\psi_{\eta_{1}\eta_{2}\eta^{\prime}_{1}\eta^{\prime}_{2}}&\sim&  \left(\vec{\Delta}_{\eta_{1}X}\cdot\vec{\Delta}_{\eta_{2}X}+
\vec{\Delta}_{\eta^\prime_{1}Y}\cdot\vec{\Delta}_{\eta^\prime_{2}Y}\right), \\
\phi_{\eta_{1}\eta_{2}\eta^{\prime}_{1}\eta^{\prime}_{2}}&\sim&  \left(\vec{\Delta}_{\eta_{1}X}
\cdot\vec{\Delta}_{\eta_{2}X}-\vec{\Delta}_{\eta^{\prime}_{1}Y}
\cdot\vec{\Delta}_{\eta^{\prime}_{2}Y}\right),
\label{eq:gen_phi}
\eea
so that 
\bea 
\vec{\Delta}_{\eta_{1}X}\cdot\vec{\Delta}_{\eta_{2}X} &\sim& 
\psi_{\eta_{1}\eta_{2}\eta^{\prime}_{1} \eta^{\prime}_{2}} + \phi_{\eta_{1}\eta_{2}\eta^{\prime}_{1}\eta^{\prime}_{2}} \\
\vec{\Delta}_{\eta_{1}Y} \cdot \vec{\Delta}_{\eta_{2}Y} &\sim& 
\psi_{\eta^{\prime}_{1}\eta^{\prime}_{2} \eta_{1}\eta_{2}} - \phi_{\eta^{\prime}_{1}\eta^{\prime}_{2}\eta_{1}\eta_{2}}. 
\eea
For simplicity we choose $\eta^\prime_{1} \eta^{\prime}_2 = \eta_{1} \eta_2$ in 
the  the $\psi, \phi$ definitions Eq.s~\pref{eq:gen_psi}-\pref{eq:gen_phi}. 
Substituting these definitions for writing the quartic terms of 
Eq.~\pref{eq:Sfinal} and reorganizing the various contributions one recovers the 
expression quoted in the main text for the effective action, 
Eq.~\pref{eq:Seff}. 

\section{QUANTITATIVE RESULTS FOR THE CONTINUUM LIMIT OF THE TWO--ORBITAL MODEL} 
\label{sect:appB}

We consider a tight-binding model for the 
$d_{yz}$ and $d_{zx}$ Fe orbitals in an As tetrahedral environment 
\cite{nosotrasnjp09, graser09}, which is the minimal model to illustrate 
how symmetry and topology affect the S-N picture. Taking into account 
the symmetries of the orbitals and the Fe square lattice, the hoppings fulfill the 
following relations: 
$t_{1}=t^{x}_{yz \, yz}=t^{y}_{zx \, zx}$, $t_{2}=t^{y}_{yz \, yz}=t^{x}_{zx \, zx}$, 
$t_{3}=t'_{zx \, zx}=t'_{yz \, yz}$, $t_{4}=t'_{zx \, yz}=-t'_{yz \, zx}$, and 
$t^{x}_{zx \, yz}=t^{y}_{zx \, yz}=0$ \cite{nosotrasnjp09, graser09}. 
The subscript indicates orbitals, the superscript $'$ indicates second nearest 
neighbors and the superscript $x/y$ the $x/y$-direction. The singular $C_4$ 
symmetry between the $yz$ and $zx$ orbitals gives rise to the non-trivial 
topology of the FS as shown in the following.

We can write the Hamiltonian in the basis of the Pauli matrices with orbital 
pseudospin $\Psi^\dagger_{\mathbf{k}}=\left(c^\dagger_{\mathbf{k},yz}, c^\dagger_{\mathbf{k},zx}\right)$:
\begin{equation}
\hat{H}_{0}=\sum_{\mathbf{k}}\Psi^{+}_{\mathbf{k}}\left((h_{0}(\mathbf{k}) - \mu )\hat{\tau}_{0}+
\vec{h}(\mathbf{k})\cdot\vec{\tau}\right  )\Psi_{\mathbf{k}},\label{Ham0}
\end{equation}
where $\mu$ is the chemical potential and  
\begin{equation}
h_{0}(\mathbf{k})=-(t_{1}+t_{2})(\cos ak_{x}+\cos ak_{y})-4t_{3}\cos ak_{x}\cos ak_{y}, \nn
\end{equation}
\begin{equation}
h_{1}(\mathbf{k})=-4t_{4}\sin ak_{x}\sin ak_{y}, \nn
\end{equation}
\begin{equation}
h_{3}(\mathbf{k})=-(t_{1}-t_{2})(\cos ak_{x}-\cos ak_{y}). \nn
\end{equation}
\begin{figure}
\includegraphics[clip,width=0.43\textwidth]{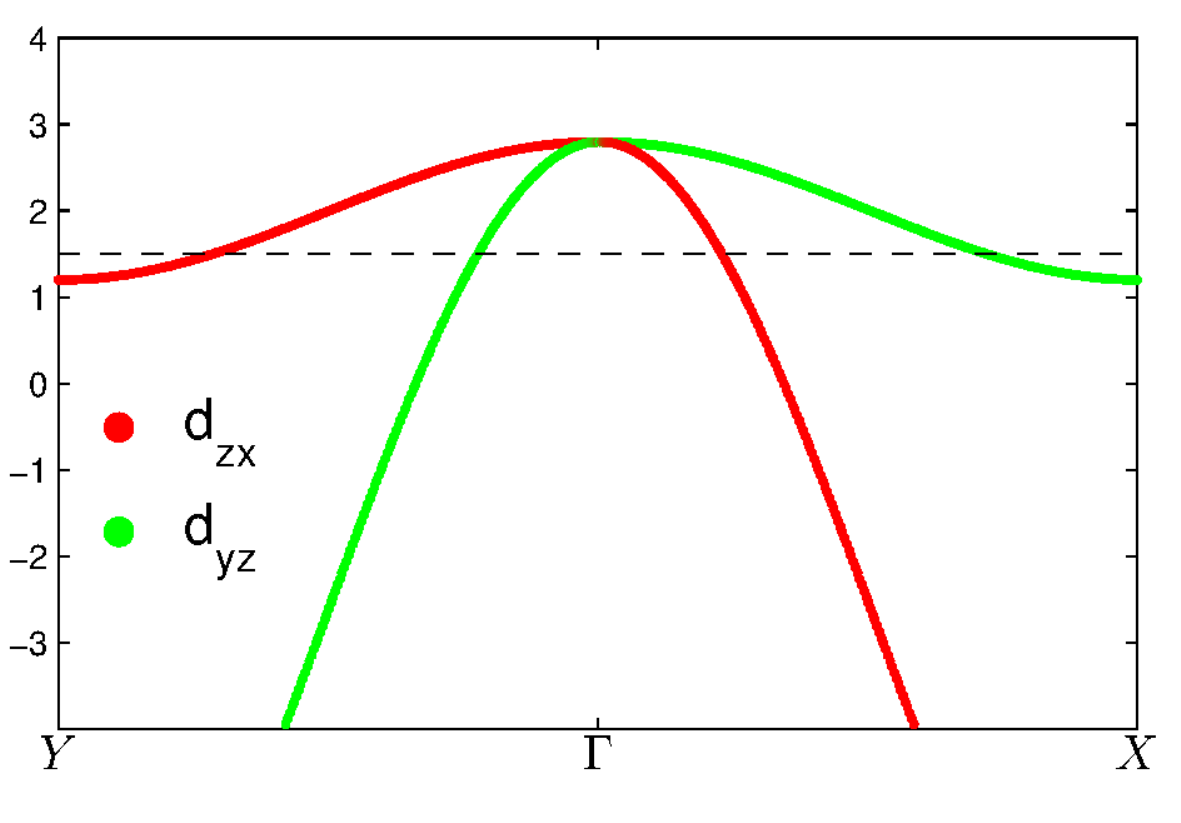}
\caption{(Color online) Band structure of the two-orbital $d_{yz}$\,--\, $d_{zx}$ model for 
the same Hamiltonian parameters used in [\onlinecite{graser09}]: 
$t_1=-1, t_2=1.3, t_3=t_4=-0.85$ and $\mu=1.85$ (dashed line) in units of $|t_1|$. 
Green (red) stands for the $yz$ ($zx$) orbital weight. Both orbitals contribute
to the $\Gamma$ pockets while the $X/Y$ pocket is mostly $yz/zx$.}
\label{fig:bands}
\vspace{-0.3cm}
\end{figure} 
Diagonalizing the Hamiltonian we obtain two bands $E^\pm({\bf k})=h_0({\bf 
k}) - \mu \pm \sqrt{h_1^2({\bf k})+h_3^2({\bf k})}$. A generic band structure 
closed to the Fermi level showing the orbital weights is represented in 
Fig.~\ref{fig:bands}. Both orbitals contribute to the $\Gamma$ pockets while the 
$X/Y$ electron pocket has dominant $yz/zx$ component. The $\Gamma$ pocket has a 
vortex structure as it was pointed out in Ref.~\cite{dunghailee09} for the two and five orbital models of iron superconductors. We will 
identify the vortex in the continuum limit.

\begin{figure}[t]
\centering
\includegraphics[clip,width=0.41\textwidth]{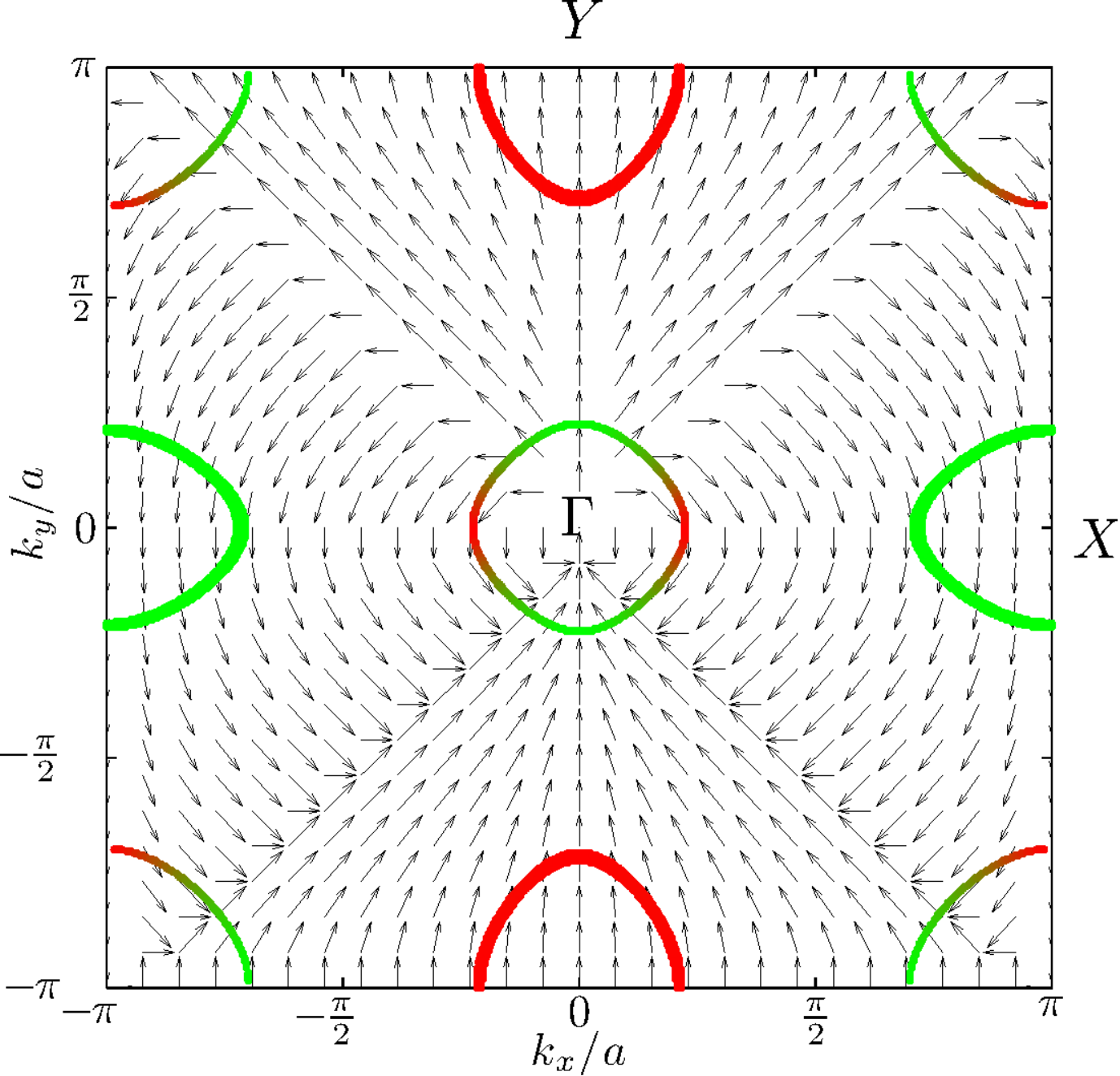}
\caption{Vortex in the orbital space: The arrows represent the vector field 
$\vec{n}=\vec{h}(\mathbf{k})/|\vec{h}(\mathbf{k})|$ in the BZ. The FSs of the 
two-orbital model, which orbital contribution is shown using the same color code 
than before ($yz$ green, $zx$ red), are superimposed. Notice the vortex around 
the $\Gamma$ point. The vector field around the Fermi surface can be identified in the continuum limit 
$\vec{n} \sim \vec{n}_\Gamma= (\sin(2\phi_\bk),\cos(2\phi_\bk))$, 
while around the $X/Y$ pockets $\vec{n} \sim \vec{n}_{X/Y} = \pm (0,1)$. 
We used the same Hamiltonian parameters of Fig.~\ref{fig:bands}.}
\label{fig:vortex}
\end{figure}

We are interested in a low energy description of the model i.e. in the behavior 
of $\hat{H}_{0}$ around the $\Gamma$, $Y$ and $X$ points. 
In the continuum limit we expand to lowest order. Around the $\Gamma$ point we 
get: 
$$h_{0,\Gamma}(\mathbf{k})\simeq 
\varepsilon_{\Gamma}+\alpha_{\Gamma}\mathbf{k}^{2}, \ \ 
\ \ h_{1,\Gamma}(\mathbf{k})\simeq-2 \, c\, k_{x}k_{y},$$ 
and 
$$h_{3,\Gamma}(\mathbf{k})\simeq b \, (k^{2}_{x}-k^{2}_{y}).$$ 
This dispersion relation has vorticity equal to two (see Fig.~\ref{fig:vortex}) 
as it has been pointed out in \cite{dunghailee09}. For simplicity we assume 
$4t_{1}\simeq t_{1}-t_{2} <0 $, so that $b\simeq c > 0 $ and around $\Gamma$ the 
dispersion relation is isotropic. 
Around the $X/Y$ point we get instead 
$$h_{0,X/Y}(\mathbf{k})\simeq \varepsilon_{X/Y}+ \alpha_{x/y}k^{2}_{y}+\alpha_{y/x}k^{2}_{y}, $$ 
$$h_{1,X/Y}(\mathbf{k})\simeq-2c k_{x}k_{y}, \ \ \ \ h_{3,X/Y}(\mathbf{k})\simeq \mp d. $$ 
Where the $\alpha_{x/y}$ coefficients show the opposite ellipticity of the e-pockets. 
All the coefficients $\varepsilon_{\Gamma},\varepsilon_{X},\varepsilon_{Y}, \alpha_\Gamma, \alpha_{x/y}$ and $b,c,d$ are 
functions of the hopping coefficients whose precise dependence is not 
important at this point.

In general, the Green's function associated to the Hamiltonian 
(\ref{Ham0}) can be written as
\begin{equation}
G(\omega,\mathbf{k})=\sum_{s=\pm1}\frac{1}{2}
\frac{\tau_{0}+s\vec{\tau}\cdot\vec{n}(\mathbf{k})}{\omega-h_{0}(\mathbf{k})-s |h(\mathbf{k})|}.
\label{eq:green}
\end{equation}
where we fixed for simplicity $\mu = 0$. The vector $\vec{n}(\mathbf{k})$ is defined as 
$\vec{h}(\mathbf{k})/|\vec{h}(\mathbf{k})|$, and $s=\pm1$ labels the conduction ($s=1$) 
and valence ($s=-1$) bands. 
We can particularize this Green's function to each pocket. The vectors $\vec{n}_m$ 
for $m= \Gamma, X, Y$ are obtained by definitions using the low energy 
expansions $\vec{h}_{m}$. Within this approximation the vector field for the 
$\Gamma$ pocket is given by 
$\vec{n}_{\Gamma}(\mathbf{k})=(\sin 2\theta_\bk, \cos 2\theta_\bk)$ where 
we use $\theta_\bk=\arctan k_y/k_x$. The vortex around $\Gamma$ 
described by $\vec{n}_\Gamma$ is represented in Fig.~\ref{fig:vortex}. For 
the $X/Y$ e-pockets we find instead $\vec{n}_{X}=-\vec{n}_{Y}=(0,1)$ that 
physically means that the orbital weight of the pocket $X/Y$ is only from the 
$d_{yz}/d_{zx}$ orbital. Using that the $\Gamma$ point belongs to the valence 
band ($s=-1$) and the $X,Y$ pockets belong to the conduction ($s=1$) band we 
have 
\begin{subequations}
\begin{equation}
\hat{G}_{\Gamma}(\omega,\mathbf{k})=\frac{1}{2}\frac{\hat{\tau_{0}}-\vec{\tau}
\cdot\vec{n}_{\Gamma}(\mathbf{k})}{\omega-\varepsilon_{\Gamma}(\mathbf{k})},
\end{equation}
\begin{equation}
\hat{G}_{X/Y}(\omega,\mathbf{k})=\frac{1}{2}\frac{\hat{\tau}_{0}+\vec{\tau}\cdot\vec{n}_{X/Y}
(\mathbf{k})}{\omega-\varepsilon_{X/Y}(\mathbf{k})}.
\end{equation}
\label{eq:greens}
\end{subequations}
By using the $\vec{n}_l$ definitions in Eq.s~(\ref{eq:greens}) we can 
explicitly write down the expressions for the orbital components of the 
Green's function of each pocket. It is easy to verify that for the $X/Y$ pockets 
all the orbitals components vanish except the $yz/zx$ one 
\begin{equation}
{G}^{11/22}_{X/Y}(\omega,\mathbf{k}) = (\omega-\varepsilon_{X/Y}(\mathbf{k}))^{-1} \equiv G_{X/Y} (\omega,\mathbf{k})
\label{eq:green_XY}
\end{equation}
using $yz=1, zx=2$, while for the $\Gamma$ pockets we have
\begin{eqnarray}
{G}^{11}_{\Gamma}(\omega,\mathbf{k}) &=& \frac{1 - \cos 2\theta_{\mathbf{k}}}{\omega-\varepsilon_{\Gamma}(\mathbf{k})} 
\equiv \sin^2 \theta_{\mathbf{k}} \, G_{\Gamma} (\omega,\mathbf{k}) \\
\label{eq:green_G11}
{G}^{12}_{\Gamma}(\omega,\mathbf{k}) &=& \frac{\sin 2\theta_{\mathbf{k}}}{\omega-\varepsilon_{\Gamma}(\mathbf{k})} 
\equiv \sin 2 \theta_{\mathbf{k}} \, G_{\Gamma} (\omega,\mathbf{k}) \\
\label{eq:green_G12}
{G}^{22}_{\Gamma}(\omega,\mathbf{k}) &=& \frac{1 + \cos 2\theta_{\mathbf{k}}}{\omega-\varepsilon_{\Gamma}(\mathbf{k})} 
\equiv\cos^2 \theta_{\mathbf{k}} \, G_{\Gamma} (\omega,\mathbf{k})
\label{eq:green_G22}
\end{eqnarray}
with ${G}^{21}_{\Gamma}={G}^{12}_{\Gamma}$ and
$G_{m}(\omega,\mathbf{k}) = (i \o_n - \varepsilon_m (\mathbf{k}))^{-1}$.
In the numerator of each Green's function it is 
encoded the orbital content of each pocket. 

The dispersion relations within this approximation can be expressed by 
$\varepsilon_{\Gamma}(\bk)\simeq 
\varepsilon_{0}-\frac{\mathbf{k}^{2}}{2m_{\Gamma}}$, 
$\varepsilon_{X/Y}(\bk)\simeq -\varepsilon_{0} + \frac{k^{2}_{x}}{2m_{x/y}} + 
\frac{k^{2}_{y}}{2m_{y/x}}$, $\varepsilon_0$ is the offset energy. 
To better compare with previous results we treat the ellipticity perturbatively 
analogously to what is discussed in Ref.~\cite{schmalianprb12}. We assume the 
isotropic mass for the h-pocket as $m_\Gamma = m$ and we use the parameter 
$\delta m $ encoding the ellipticity, such that the anisotropic e-pocket mass read as $ 1/m_{x/y} = (1 \pm 
\delta m)/ m$.
Near the Fermi energy and for small ellipticity the dispersions can be 
approximate by $\varepsilon_\Gamma = - \varepsilon$ and 
$\varepsilon_{X/Y} = \varepsilon \pm \varepsilon_0 \delta m \cos 2 \theta_{\mathbf{k}}$. 

\paragraph{Magnetic Susceptibility in the Nematic Phase}

In general the spin susceptibility depend on four orbital indices but in the 
spin-nematic scenario the spin-susceptibility, Eq.s~(6-7) of the main text, 
depends just on two orbitals indices \cite{note_orbmagn}. 
In the continuum limit the magnetic $(\pi, 0)$ bubble is given by 
\begin{equation}
\Pi^{\eta\eta'}_X (q)=\frac{1}{2}\sum_{k} G^{\eta\eta'}_{\Gamma}(k)\, G^{\eta'\eta}_{X}(k+q),
\end{equation}
where there is no summation in $\eta,\eta'$ and we compact the notation using 
the quadrivectors $k\equiv (i\o_n, \bk)$,  $q \equiv(i\O_m, \bq)$ and 
$\sum_k \equiv T\sum_{\omega_{n}}\int \frac{d^{2}\mathbf{k}}{4\pi^{2}}$.
By replacing the Green's functions by Eq.~(\ref{eq:green_XY}-\ref{eq:green_G22}) 
we obtain that the only finite component is the $\Pi^{11}_X$ 
\begin{equation}
\Pi^{11}_X(q)=\frac{1}{2}\sum_{k} \, \sin^2 \theta_{\mathbf{k}} \ G_{\Gamma}(k)G_{X}(k+q).
\label{eq:pi11_complete}
\end{equation}
By comparison of the above equation with the expression for the bubble 
defined in terms of the orbital weights $\o^{\eta_1}_{\Gamma X} 
\o^{\eta_1}_{\Gamma X}$, Eq.~(7) of the main text, we 
identify 
\begin{equation}
\o_{\G X}^{1} \o_{\G X}^{1} = \sin^2{\theta_\bk}, 
\ \ \ \ \ \o_{\G X}^{1} \o_{\G X}^{2} = \o_{\G X}^{2} \o_{\G X}^{2} = 0 .
\label{eq:wX_def}
\end{equation}
By analogous analysis of the $(0,\pi)$ bubble $\Pi^{\eta \eta^\prime}_Y$ we obtain 
\begin{equation}
\o_{\G Y}^{2}\o_{\G Y}^{2}=\cos^2{\theta_\bk}, 
\ \ \ \ \ \o_{\G Y}^{1} \o_{\G Y}^{2} = \o_{\G Y}^{1} \o_{\G Y}^{1} = 0 .
\label{eq:wY_def}
\end{equation}

Let us focus now on the hydrodynamic limit of the static 
$(\pi, 0)$ bubble Eq.~(\ref{eq:pi11_complete}) in the nematic phase
\begin{equation} 
\Pi^{11}_{X}(0,\mathbf{q})=\Pi^{11}_X(0)+ c_X(\theta_{\mathbf{q}})\mathbf{q}^{2}+\mathcal{O}(\mathbf{q}^{3}). 
\end{equation} 
To study the ${\mathbf q}$--dependent part, we have to expand up to $\mathbf 
q^2$ the $G_{X}(k+q)$ Green's function in Eq.~(\ref{eq:pi11_complete}) also 
taking into account the ellipticity of the $X$ pockets. 
The constant part and the momentum dependent one can be computed explicitly 
separating the momentum integral in the one over the angular variable 
$\theta_{\mathbf{k}}$ and the other over the modulus $\int dk^2 \simeq N_{F}\int 
d\varepsilon$ with $N_{F}$ being the density of state at the Fermi level.

The constant part is negative and at perfect nesting diverges logarithmically as
$ \Pi^{11}_X(0) \simeq - N_{F}\log(\Lambda/2T)/(4 \pi) + ...$,
with $\Lambda$ upper cut-off for the low-energy theory. The term is the 
standard logarithm appearing in the problem of the antiferromagnetic 
instability. 

The $\mathbf{q}$-dependent part can be written as 
$c_X(\theta_{\mathbf{q}}){\mathbf{q}}^{2}=c_X^x q_x^{2}+ c_X^y q_y^{2} $ where 
\begin{equation}
c_X^x =  -\frac{v_F^2 N_{F}}{256 \pi T^2}(1+\delta m), \ \  
c_X^y =  -\frac{v_F^2 N_{F}}{256 \pi T^2}(3-7\delta m).
\label{eq:stiffness}
\end{equation}
Thus we found that the stiffness of the magnetic bubble is anisotropic already 
at the level of zero ellipticity $\delta m=0$ due to the angular modulation 
$\sin^2 \theta_{\mathbf{k}}$  introduced by the vortex in 
Eq.~(\ref{eq:pi11_complete}). Notice that the finite term and the one proportional 
to $\delta m $ have a competing role in making the magnetic stiffness 
anisotropic.
Eq.~(\ref{eq:pi11_complete}) reproduces exactly the magnetic bubble obtained via 
orbital-less approach \cite{schmalianprb12} once eliminated $\sin^2 
\theta_{\mathbf{k}}$. The momenta analysis for this case leads to:
\begin{equation}
c_X^x = -\frac{v_F^2 N_{F}}{64 \pi T^2}(1+2\delta m),  \ \ 
c_X^y =  -\frac{v_F^2 N_{F}}{64 \pi T^2}(1-2\delta m),
\end{equation}
and the anisotropic stiffness is found only at order $\delta 
m$. In addition, the dependence on $\delta m$ of the $c_X^{x/y}$ 
coefficients is different with respect to the one obtain 
retaining the orbital information, Eq.s~(\ref{eq:stiffness}) highlight that the vortex 
also affects the terms accounting for the ellipticity. 

Notice that in the paramagnetic phase the tetragonal 
symmetry is respected since the anisotropic properties found for the $\Pi_X$ 
hold for the $\Pi_Y$ having into account the orbital exchange $yz/xz$ i.e. $c_Y^x 
\equiv c_X^y$ and $c_Y^y \equiv c_X^x$. Once is entered in the nematic phase the two 
modes are no longer equivalent and the momentum dependence of the spin 
fluctuations is actually anisotropic as we discussed above. 

\paragraph{Quartic Coefficients} 

We start directly from the definition of the orbital nematic coefficients $\hat{g}$ Eq.~(5c) of the main text. 
By using the expressions Eq.s~(\ref{eq:wX_def}-\ref{eq:wY_def}) for the orbital weights we have 
\begin{subequations}
\begin{equation}
g_{1111}= - \frac{1}{2}\sum_{k}\, \sin^{4}(\theta_{\mathbf{k}}) G^{2}_{\Gamma}(k) G^{2}_{X}(k),
\end{equation}
\begin{equation}
g_{1122}=\frac{1}{2}\sum_{k} \,\sin^{2}(\theta_\bk) \cos^{2}(\theta_{\bk}) G^{2}_{\Gamma}(k)G_{X}(k) G_{Y}(k).
\end{equation}
\end{subequations}
Expanding $G_{X/Y}$ in powers of $\delta m$ and computing explicitly the integrals we obtain 
\begin{equation}
g_{11\eta\eta}=\frac{N_F}{\pi T^2}\;a_\eta \bigg( 1+b_\eta\;\frac{\epsilon_0^2}{T^2} \delta m^{2}\bigg)\; ; 
\end{equation}
with $a_\eta$ and $b_\eta$ ($\eta=1,2$) numerical factors: $- a_1=3\, a_2=3/2^9$ and $b_1=7\, b_2=7/2^6$. 
It holds $g_{1111} = g_{2222}, g_{1122}=g_{2211}$. The tensor elements of the 
nematic coupling are finite also assuming circular pockets (i.e. $\delta m=0$) 
while the first finite contribution in $\delta m$ appears at the second order. 
All the others quartic coefficients $\hat{u}$, $\hat{v}$ are related to the 
above ones by angular symmetry. It holds: $u_{2222}=u_{1111}= - g_{1111}$, 
$u_{1122}=u_{2211} = g_{1122}$, $v_{1111}= - v_{2222}= - g_{1111}$, 
$v_{1122}= v_{2211}= - g_{1122}$ while all the others 
components are zero.

\bibliography{pnictides_nematic}

\begin{thebibliography}{76}%
\makeatletter
\providecommand \@ifxundefined [1]{%
 \@ifx{#1\undefined}
}%
\providecommand \@ifnum [1]{%
 \ifnum #1\expandafter \@firstoftwo
 \else \expandafter \@secondoftwo
 \fi
}%
\providecommand \@ifx [1]{%
 \ifx #1\expandafter \@firstoftwo
 \else \expandafter \@secondoftwo
 \fi
}%
\providecommand \natexlab [1]{#1}%
\providecommand \enquote  [1]{``#1''}%
\providecommand \bibnamefont  [1]{#1}%
\providecommand \bibfnamefont [1]{#1}%
\providecommand \citenamefont [1]{#1}%
\providecommand \href@noop [0]{\@secondoftwo}%
\providecommand \href [0]{\begingroup \@sanitize@url \@href}%
\providecommand \@href[1]{\@@startlink{#1}\@@href}%
\providecommand \@@href[1]{\endgroup#1\@@endlink}%
\providecommand \@sanitize@url [0]{\catcode `\\12\catcode `\$12\catcode
  `\&12\catcode `\#12\catcode `\^12\catcode `\_12\catcode `\%12\relax}%
\providecommand \@@startlink[1]{}%
\providecommand \@@endlink[0]{}%
\providecommand \url  [0]{\begingroup\@sanitize@url \@url }%
\providecommand \@url [1]{\endgroup\@href {#1}{\urlprefix }}%
\providecommand \urlprefix  [0]{URL }%
\providecommand \Eprint [0]{\href }%
\providecommand \doibase [0]{http://dx.doi.org/}%
\providecommand \selectlanguage [0]{\@gobble}%
\providecommand \bibinfo  [0]{\@secondoftwo}%
\providecommand \bibfield  [0]{\@secondoftwo}%
\providecommand \translation [1]{[#1]}%
\providecommand \BibitemOpen [0]{}%
\providecommand \bibitemStop [0]{}%
\providecommand \bibitemNoStop [0]{.\EOS\space}%
\providecommand \EOS [0]{\spacefactor3000\relax}%
\providecommand \BibitemShut  [1]{\csname bibitem#1\endcsname}%
\let\auto@bib@innerbib\@empty
\bibitem [{\citenamefont {Tanatar}\ \emph {et~al.}(2010)\citenamefont
  {Tanatar}, \citenamefont {Blomberg}, \citenamefont {Kreyssig}, \citenamefont
  {Kim}, \citenamefont {Ni}, \citenamefont {Thaler}, \citenamefont {Bud'ko},
  \citenamefont {Canfield}, \citenamefont {Goldman}, \citenamefont {Mazin},\
  and\ \citenamefont {Prozorov}}]{mazin10}%
  \BibitemOpen
  \bibfield  {author} {\bibinfo {author} {\bibfnamefont {M.~A.}\ \bibnamefont
  {Tanatar}}, \bibinfo {author} {\bibfnamefont {E.~C.}\ \bibnamefont
  {Blomberg}}, \bibinfo {author} {\bibfnamefont {A.}~\bibnamefont {Kreyssig}},
  \bibinfo {author} {\bibfnamefont {M.~G.}\ \bibnamefont {Kim}}, \bibinfo
  {author} {\bibfnamefont {N.}~\bibnamefont {Ni}}, \bibinfo {author}
  {\bibfnamefont {A.}~\bibnamefont {Thaler}}, \bibinfo {author} {\bibfnamefont
  {S.~L.}\ \bibnamefont {Bud'ko}}, \bibinfo {author} {\bibfnamefont {P.~C.}\
  \bibnamefont {Canfield}}, \bibinfo {author} {\bibfnamefont {A.~I.}\
  \bibnamefont {Goldman}}, \bibinfo {author} {\bibfnamefont {I.~I.}\
  \bibnamefont {Mazin}}, \ and\ \bibinfo {author} {\bibfnamefont
  {R.}~\bibnamefont {Prozorov}},\ }\href {\doibase 10.1103/PhysRevB.81.184508}
  {\bibfield  {journal} {\bibinfo  {journal} {Phys. Rev. B}\ }\textbf {\bibinfo
  {volume} {81}},\ \bibinfo {pages} {184508} (\bibinfo {year}
  {2010})}\BibitemShut {NoStop}%
\bibitem [{\citenamefont {Chu}\ \emph {et~al.}(2010)\citenamefont {Chu},
  \citenamefont {Analytis}, \citenamefont {De~Greve}, \citenamefont {McMahon},
  \citenamefont {Islam}, \citenamefont {Yamamoto},\ and\ \citenamefont
  {Fisher}}]{chuscience2010}%
  \BibitemOpen
  \bibfield  {author} {\bibinfo {author} {\bibfnamefont {J.-H.}\ \bibnamefont
  {Chu}}, \bibinfo {author} {\bibfnamefont {J.~G.}\ \bibnamefont {Analytis}},
  \bibinfo {author} {\bibfnamefont {K.}~\bibnamefont {De~Greve}}, \bibinfo
  {author} {\bibfnamefont {P.~L.}\ \bibnamefont {McMahon}}, \bibinfo {author}
  {\bibfnamefont {Z.}~\bibnamefont {Islam}}, \bibinfo {author} {\bibfnamefont
  {Y.}~\bibnamefont {Yamamoto}}, \ and\ \bibinfo {author} {\bibfnamefont
  {I.~R.}\ \bibnamefont {Fisher}},\ }\href@noop {} {\bibfield  {journal}
  {\bibinfo  {journal} {Science}\ }\textbf {\bibinfo {volume} {329}},\ \bibinfo
  {pages} {824} (\bibinfo {year} {2010})}\BibitemShut {NoStop}%
\bibitem [{\citenamefont {Fisher}\ \emph {et~al.}(2011)\citenamefont {Fisher},
  \citenamefont {Degiorgi},\ and\ \citenamefont {Shen}}]{fisher2011}%
  \BibitemOpen
  \bibfield  {author} {\bibinfo {author} {\bibfnamefont {I.~R.}\ \bibnamefont
  {Fisher}}, \bibinfo {author} {\bibfnamefont {L.}~\bibnamefont {Degiorgi}}, \
  and\ \bibinfo {author} {\bibfnamefont {Z.~X.}\ \bibnamefont {Shen}},\
  }\href@noop {} {\bibfield  {journal} {\bibinfo  {journal} {Rep. Prog. Phys.}\
  }\textbf {\bibinfo {volume} {74}},\ \bibinfo {pages} {124506} (\bibinfo
  {year} {2011})}\BibitemShut {NoStop}%
\bibitem [{\citenamefont {Ying}\ \emph {et~al.}(2011)\citenamefont {Ying},
  \citenamefont {Wang}, \citenamefont {Wu}, \citenamefont {Xiang},
  \citenamefont {Liu}, \citenamefont {Yan}, \citenamefont {Wang}, \citenamefont
  {Zhang}, \citenamefont {Ye}, \citenamefont {Cheng}, \citenamefont {Hu},\ and\
  \citenamefont {Chen}}]{yingprl11}%
  \BibitemOpen
  \bibfield  {author} {\bibinfo {author} {\bibfnamefont {J.~J.}\ \bibnamefont
  {Ying}}, \bibinfo {author} {\bibfnamefont {X.~F.}\ \bibnamefont {Wang}},
  \bibinfo {author} {\bibfnamefont {T.}~\bibnamefont {Wu}}, \bibinfo {author}
  {\bibfnamefont {Z.~J.}\ \bibnamefont {Xiang}}, \bibinfo {author}
  {\bibfnamefont {R.~H.}\ \bibnamefont {Liu}}, \bibinfo {author} {\bibfnamefont
  {Y.~J.}\ \bibnamefont {Yan}}, \bibinfo {author} {\bibfnamefont {A.~F.}\
  \bibnamefont {Wang}}, \bibinfo {author} {\bibfnamefont {M.}~\bibnamefont
  {Zhang}}, \bibinfo {author} {\bibfnamefont {G.~J.}\ \bibnamefont {Ye}},
  \bibinfo {author} {\bibfnamefont {P.}~\bibnamefont {Cheng}}, \bibinfo
  {author} {\bibfnamefont {J.~P.}\ \bibnamefont {Hu}}, \ and\ \bibinfo {author}
  {\bibfnamefont {X.~H.}\ \bibnamefont {Chen}},\ }\href@noop {} {\bibfield
  {journal} {\bibinfo  {journal} {Phys. Rev. Lett.}\ }\textbf {\bibinfo
  {volume} {107}},\ \bibinfo {pages} {067001} (\bibinfo {year}
  {2011})}\BibitemShut {NoStop}%
\bibitem [{\citenamefont {Chu}\ \emph {et~al.}(2012)\citenamefont {Chu},
  \citenamefont {Kuo}, \citenamefont {Analytis},\ and\ \citenamefont
  {Fisher}}]{chuscience2012}%
  \BibitemOpen
  \bibfield  {author} {\bibinfo {author} {\bibfnamefont {J.-H.}\ \bibnamefont
  {Chu}}, \bibinfo {author} {\bibfnamefont {H.-H.}\ \bibnamefont {Kuo}},
  \bibinfo {author} {\bibfnamefont {J.~G.}\ \bibnamefont {Analytis}}, \ and\
  \bibinfo {author} {\bibfnamefont {I.~R.}\ \bibnamefont {Fisher}},\
  }\href@noop {} {\bibfield  {journal} {\bibinfo  {journal} {Science}\ }\textbf
  {\bibinfo {volume} {337}},\ \bibinfo {pages} {710} (\bibinfo {year}
  {2012})}\BibitemShut {NoStop}%
\bibitem [{\citenamefont {Kuo}\ and\ \citenamefont
  {Fisher}(2014)}]{fisherprl14}%
  \BibitemOpen
  \bibfield  {author} {\bibinfo {author} {\bibfnamefont {H.-H.}\ \bibnamefont
  {Kuo}}\ and\ \bibinfo {author} {\bibfnamefont {I.~R.}\ \bibnamefont
  {Fisher}},\ }\href@noop {} {\bibfield  {journal} {\bibinfo  {journal} {Phys.
  Rev. Lett.}\ }\textbf {\bibinfo {volume} {112}},\ \bibinfo {pages} {227001}
  (\bibinfo {year} {2014})}\BibitemShut {NoStop}%
\bibitem [{\citenamefont {Jiang}\ \emph {et~al.}(2013)\citenamefont {Jiang},
  \citenamefont {Jeevan}, \citenamefont {Dong},\ and\ \citenamefont
  {Gegenwart}}]{jiangprl13}%
  \BibitemOpen
  \bibfield  {author} {\bibinfo {author} {\bibfnamefont {S.}~\bibnamefont
  {Jiang}}, \bibinfo {author} {\bibfnamefont {H.~S.}\ \bibnamefont {Jeevan}},
  \bibinfo {author} {\bibfnamefont {J.}~\bibnamefont {Dong}}, \ and\ \bibinfo
  {author} {\bibfnamefont {P.}~\bibnamefont {Gegenwart}},\ }\href@noop {}
  {\bibfield  {journal} {\bibinfo  {journal} {Phys. Rev. Lett.}\ }\textbf
  {\bibinfo {volume} {110}},\ \bibinfo {pages} {067001} (\bibinfo {year}
  {2013})}\BibitemShut {NoStop}%
\bibitem [{\citenamefont {Fernandes}\ \emph {et~al.}(2010)\citenamefont
  {Fernandes}, \citenamefont {VanBebber}, \citenamefont {Bhattacharya},
  \citenamefont {Chandra}, \citenamefont {Keppens}, \citenamefont {Mandrus},
  \citenamefont {McGuire}, \citenamefont {Sales}, \citenamefont {Sefat},\ and\
  \citenamefont {Schmalian}}]{Fernandesprl10}%
  \BibitemOpen
  \bibfield  {author} {\bibinfo {author} {\bibfnamefont {R.~M.}\ \bibnamefont
  {Fernandes}}, \bibinfo {author} {\bibfnamefont {L.~H.}\ \bibnamefont
  {VanBebber}}, \bibinfo {author} {\bibfnamefont {S.}~\bibnamefont
  {Bhattacharya}}, \bibinfo {author} {\bibfnamefont {P.}~\bibnamefont
  {Chandra}}, \bibinfo {author} {\bibfnamefont {V.}~\bibnamefont {Keppens}},
  \bibinfo {author} {\bibfnamefont {D.}~\bibnamefont {Mandrus}}, \bibinfo
  {author} {\bibfnamefont {M.~A.}\ \bibnamefont {McGuire}}, \bibinfo {author}
  {\bibfnamefont {B.~C.}\ \bibnamefont {Sales}}, \bibinfo {author}
  {\bibfnamefont {A.~S.}\ \bibnamefont {Sefat}}, \ and\ \bibinfo {author}
  {\bibfnamefont {J.}~\bibnamefont {Schmalian}},\ }\href@noop {} {\bibfield
  {journal} {\bibinfo  {journal} {Phys. Rev. Lett.}\ }\textbf {\bibinfo
  {volume} {105}},\ \bibinfo {pages} {157003} (\bibinfo {year}
  {2010})}\BibitemShut {NoStop}%
\bibitem [{\citenamefont {B{\"o}hmer}\ \emph {et~al.}(2014)\citenamefont
  {B{\"o}hmer}, \citenamefont {Burger}, \citenamefont {Hardy}, \citenamefont
  {Wolf}, \citenamefont {Schweiss}, \citenamefont {Fromknecht}, \citenamefont
  {Reinecker}, \citenamefont {Schranz},\ and\ \citenamefont
  {Meingast}}]{meingastprl04}%
  \BibitemOpen
  \bibfield  {author} {\bibinfo {author} {\bibfnamefont {A.~E.}\ \bibnamefont
  {B{\"o}hmer}}, \bibinfo {author} {\bibfnamefont {P.}~\bibnamefont {Burger}},
  \bibinfo {author} {\bibfnamefont {F.}~\bibnamefont {Hardy}}, \bibinfo
  {author} {\bibfnamefont {T.}~\bibnamefont {Wolf}}, \bibinfo {author}
  {\bibfnamefont {P.}~\bibnamefont {Schweiss}}, \bibinfo {author}
  {\bibfnamefont {R.}~\bibnamefont {Fromknecht}}, \bibinfo {author}
  {\bibfnamefont {M.}~\bibnamefont {Reinecker}}, \bibinfo {author}
  {\bibfnamefont {W.}~\bibnamefont {Schranz}}, \ and\ \bibinfo {author}
  {\bibfnamefont {C.}~\bibnamefont {Meingast}},\ }\href@noop {} {\bibfield
  {journal} {\bibinfo  {journal} {Phys. Rev. Lett.}\ }\textbf {\bibinfo
  {volume} {112}},\ \bibinfo {pages} {047001} (\bibinfo {year}
  {2014})}\BibitemShut {NoStop}%
\bibitem [{\citenamefont {Dusza}\ \emph {et~al.}(2011)\citenamefont {Dusza},
  \citenamefont {Lucarelli}, \citenamefont {Pfuner}, \citenamefont {Chu},
  \citenamefont {Fisher},\ and\ \citenamefont {Degiorgi}}]{degiorgi10}%
  \BibitemOpen
  \bibfield  {author} {\bibinfo {author} {\bibfnamefont {A.}~\bibnamefont
  {Dusza}}, \bibinfo {author} {\bibfnamefont {A.}~\bibnamefont {Lucarelli}},
  \bibinfo {author} {\bibfnamefont {F.}~\bibnamefont {Pfuner}}, \bibinfo
  {author} {\bibfnamefont {J.-H.}\ \bibnamefont {Chu}}, \bibinfo {author}
  {\bibfnamefont {I.}~\bibnamefont {Fisher}}, \ and\ \bibinfo {author}
  {\bibfnamefont {L.}~\bibnamefont {Degiorgi}},\ }\href@noop {} {\bibfield
  {journal} {\bibinfo  {journal} {Europhys. Lett.}\ }\textbf {\bibinfo {volume}
  {93}},\ \bibinfo {pages} {37002} (\bibinfo {year} {2011})}\BibitemShut
  {NoStop}%
\bibitem [{\citenamefont {Nakajima}\ \emph {et~al.}(2011)\citenamefont
  {Nakajima}, \citenamefont {Liang}, \citenamefont {Ishida}, \citenamefont
  {Tomioka}, \citenamefont {Kihou}, \citenamefont {Lee}, \citenamefont {Iyo},
  \citenamefont {Eisaki}, \citenamefont {Kakeshita}, \citenamefont {Ito},\ and\
  \citenamefont {Uchida}}]{uchida2011}%
  \BibitemOpen
  \bibfield  {author} {\bibinfo {author} {\bibfnamefont {M.}~\bibnamefont
  {Nakajima}}, \bibinfo {author} {\bibfnamefont {T.}~\bibnamefont {Liang}},
  \bibinfo {author} {\bibfnamefont {S.}~\bibnamefont {Ishida}}, \bibinfo
  {author} {\bibfnamefont {Y.}~\bibnamefont {Tomioka}}, \bibinfo {author}
  {\bibfnamefont {K.}~\bibnamefont {Kihou}}, \bibinfo {author} {\bibfnamefont
  {C.~H.}\ \bibnamefont {Lee}}, \bibinfo {author} {\bibfnamefont
  {A.}~\bibnamefont {Iyo}}, \bibinfo {author} {\bibfnamefont {H.}~\bibnamefont
  {Eisaki}}, \bibinfo {author} {\bibfnamefont {T.}~\bibnamefont {Kakeshita}},
  \bibinfo {author} {\bibfnamefont {T.}~\bibnamefont {Ito}}, \ and\ \bibinfo
  {author} {\bibfnamefont {S.}~\bibnamefont {Uchida}},\ }\href@noop {}
  {\bibfield  {journal} {\bibinfo  {journal} {PNAS}\ }\textbf {\bibinfo
  {volume} {108}},\ \bibinfo {pages} {12238} (\bibinfo {year}
  {2011})}\BibitemShut {NoStop}%
\bibitem [{\citenamefont {Dusza}\ \emph {et~al.}(2012)\citenamefont {Dusza},
  \citenamefont {Lucarelli}, \citenamefont {Sanna}, \citenamefont {Massidda},
  \citenamefont {Chu}, \citenamefont {Fisher},\ and\ \citenamefont
  {Degiorgi}}]{degiorgi2012}%
  \BibitemOpen
  \bibfield  {author} {\bibinfo {author} {\bibfnamefont {A.}~\bibnamefont
  {Dusza}}, \bibinfo {author} {\bibfnamefont {A.}~\bibnamefont {Lucarelli}},
  \bibinfo {author} {\bibfnamefont {A.}~\bibnamefont {Sanna}}, \bibinfo
  {author} {\bibfnamefont {S.}~\bibnamefont {Massidda}}, \bibinfo {author}
  {\bibfnamefont {J.-H.}\ \bibnamefont {Chu}}, \bibinfo {author} {\bibfnamefont
  {I.~R.}\ \bibnamefont {Fisher}}, \ and\ \bibinfo {author} {\bibfnamefont
  {L.}~\bibnamefont {Degiorgi}},\ }\href@noop {} {\bibfield  {journal}
  {\bibinfo  {journal} {New J. Phys.}\ }\textbf {\bibinfo {volume} {14}},\
  \bibinfo {pages} {023020} (\bibinfo {year} {2012})}\BibitemShut {NoStop}%
\bibitem [{\citenamefont {Nakajima}\ \emph {et~al.}(2012)\citenamefont
  {Nakajima}, \citenamefont {Ishida}, \citenamefont {Tomioka}, \citenamefont
  {Kihou}, \citenamefont {Lee}, \citenamefont {Iyo}, \citenamefont {Ito},
  \citenamefont {Kakeshita}, \citenamefont {Eisaki},\ and\ \citenamefont
  {Uchida}}]{nakajimaprl12}%
  \BibitemOpen
  \bibfield  {author} {\bibinfo {author} {\bibfnamefont {M.}~\bibnamefont
  {Nakajima}}, \bibinfo {author} {\bibfnamefont {S.}~\bibnamefont {Ishida}},
  \bibinfo {author} {\bibfnamefont {Y.}~\bibnamefont {Tomioka}}, \bibinfo
  {author} {\bibfnamefont {K.}~\bibnamefont {Kihou}}, \bibinfo {author}
  {\bibfnamefont {C.~H.}\ \bibnamefont {Lee}}, \bibinfo {author} {\bibfnamefont
  {A.}~\bibnamefont {Iyo}}, \bibinfo {author} {\bibfnamefont {T.}~\bibnamefont
  {Ito}}, \bibinfo {author} {\bibfnamefont {T.}~\bibnamefont {Kakeshita}},
  \bibinfo {author} {\bibfnamefont {H.}~\bibnamefont {Eisaki}}, \ and\ \bibinfo
  {author} {\bibfnamefont {S.}~\bibnamefont {Uchida}},\ }\href@noop {}
  {\bibfield  {journal} {\bibinfo  {journal} {Phys. Rev. Lett.}\ }\textbf
  {\bibinfo {volume} {109}},\ \bibinfo {pages} {217003} (\bibinfo {year}
  {2012})}\BibitemShut {NoStop}%
\bibitem [{\citenamefont {Arham}\ \emph {et~al.}(2012)\citenamefont {Arham},
  \citenamefont {Hunt}, \citenamefont {Park}, \citenamefont {Gillett},
  \citenamefont {Das}, \citenamefont {Sebastian}, \citenamefont {Xu},
  \citenamefont {Wen}, \citenamefont {Lin}, \citenamefont {Li}, \citenamefont
  {Gu}, \citenamefont {Thaler}, \citenamefont {Ran}, \citenamefont {Bud'ko},
  \citenamefont {Canfield}, \citenamefont {Chung}, \citenamefont {Kanatzidis},\
  and\ \citenamefont {Greene}}]{laura-greenprb12}%
  \BibitemOpen
  \bibfield  {author} {\bibinfo {author} {\bibfnamefont {H.~Z.}\ \bibnamefont
  {Arham}}, \bibinfo {author} {\bibfnamefont {C.~R.}\ \bibnamefont {Hunt}},
  \bibinfo {author} {\bibfnamefont {W.~K.}\ \bibnamefont {Park}}, \bibinfo
  {author} {\bibfnamefont {J.}~\bibnamefont {Gillett}}, \bibinfo {author}
  {\bibfnamefont {S.~D.}\ \bibnamefont {Das}}, \bibinfo {author} {\bibfnamefont
  {S.~E.}\ \bibnamefont {Sebastian}}, \bibinfo {author} {\bibfnamefont {Z.~J.}\
  \bibnamefont {Xu}}, \bibinfo {author} {\bibfnamefont {J.~S.}\ \bibnamefont
  {Wen}}, \bibinfo {author} {\bibfnamefont {Z.~W.}\ \bibnamefont {Lin}},
  \bibinfo {author} {\bibfnamefont {Q.}~\bibnamefont {Li}}, \bibinfo {author}
  {\bibfnamefont {G.}~\bibnamefont {Gu}}, \bibinfo {author} {\bibfnamefont
  {A.}~\bibnamefont {Thaler}}, \bibinfo {author} {\bibfnamefont
  {S.}~\bibnamefont {Ran}}, \bibinfo {author} {\bibfnamefont {S.~L.}\
  \bibnamefont {Bud'ko}}, \bibinfo {author} {\bibfnamefont {P.~C.}\
  \bibnamefont {Canfield}}, \bibinfo {author} {\bibfnamefont {D.~Y.}\
  \bibnamefont {Chung}}, \bibinfo {author} {\bibfnamefont {M.~G.}\ \bibnamefont
  {Kanatzidis}}, \ and\ \bibinfo {author} {\bibfnamefont {L.~H.}\ \bibnamefont
  {Greene}},\ }\href@noop {} {\bibfield  {journal} {\bibinfo  {journal} {Phys.
  Rev. B}\ }\textbf {\bibinfo {volume} {85}},\ \bibinfo {pages} {214515}
  (\bibinfo {year} {2012})}\BibitemShut {NoStop}%
\bibitem [{\citenamefont {Blomberg}\ \emph {et~al.}(2013)\citenamefont
  {Blomberg}, \citenamefont {Tanatar}, \citenamefont {Fernandes}, \citenamefont
  {Mazin}, \citenamefont {Shen}, \citenamefont {Wen}, \citenamefont {Johannes},
  \citenamefont {Schmalian},\ and\ \citenamefont
  {Prozorov}}]{prozorovnatcomm13}%
  \BibitemOpen
  \bibfield  {author} {\bibinfo {author} {\bibfnamefont {E.~C.}\ \bibnamefont
  {Blomberg}}, \bibinfo {author} {\bibfnamefont {M.~A.}\ \bibnamefont
  {Tanatar}}, \bibinfo {author} {\bibfnamefont {R.~M.}\ \bibnamefont
  {Fernandes}}, \bibinfo {author} {\bibfnamefont {I.~I.}\ \bibnamefont
  {Mazin}}, \bibinfo {author} {\bibfnamefont {B.}~\bibnamefont {Shen}},
  \bibinfo {author} {\bibfnamefont {H.-H.}\ \bibnamefont {Wen}}, \bibinfo
  {author} {\bibfnamefont {M.~D.}\ \bibnamefont {Johannes}}, \bibinfo {author}
  {\bibfnamefont {J.}~\bibnamefont {Schmalian}}, \ and\ \bibinfo {author}
  {\bibfnamefont {R.}~\bibnamefont {Prozorov}},\ }\href@noop {} {\bibfield
  {journal} {\bibinfo  {journal} {Nat Commun}\ }\textbf {\bibinfo {volume}
  {4}},\ \bibinfo {pages} {1914} (\bibinfo {year} {2013})}\BibitemShut
  {NoStop}%
\bibitem [{\citenamefont {Mirri}\ \emph {et~al.}(2014)\citenamefont {Mirri},
  \citenamefont {Dusza}, \citenamefont {Bastelberger}, \citenamefont {Chu},
  \citenamefont {Kuo}, \citenamefont {Fisher},\ and\ \citenamefont
  {Degiorgi}}]{degiorgiprb14}%
  \BibitemOpen
  \bibfield  {author} {\bibinfo {author} {\bibfnamefont {C.}~\bibnamefont
  {Mirri}}, \bibinfo {author} {\bibfnamefont {A.}~\bibnamefont {Dusza}},
  \bibinfo {author} {\bibfnamefont {S.}~\bibnamefont {Bastelberger}}, \bibinfo
  {author} {\bibfnamefont {J.-H.}\ \bibnamefont {Chu}}, \bibinfo {author}
  {\bibfnamefont {H.-H.}\ \bibnamefont {Kuo}}, \bibinfo {author} {\bibfnamefont
  {I.~R.}\ \bibnamefont {Fisher}}, \ and\ \bibinfo {author} {\bibfnamefont
  {L.}~\bibnamefont {Degiorgi}},\ }\href@noop {} {\bibfield  {journal}
  {\bibinfo  {journal} {Phys. Rev. B}\ }\textbf {\bibinfo {volume} {89}},\
  \bibinfo {pages} {060501} (\bibinfo {year} {2014})}\BibitemShut {NoStop}%
\bibitem [{\citenamefont {Gallais}\ \emph {et~al.}(2013)\citenamefont
  {Gallais}, \citenamefont {Fernandes}, \citenamefont {Paul}, \citenamefont
  {Chauvi{\`e}re}, \citenamefont {Yang}, \citenamefont {M{\'e}asson},
  \citenamefont {Cazayous}, \citenamefont {Sacuto}, \citenamefont {Colson},\
  and\ \citenamefont {Forget}}]{gallaisprl13}%
  \BibitemOpen
  \bibfield  {author} {\bibinfo {author} {\bibfnamefont {Y.}~\bibnamefont
  {Gallais}}, \bibinfo {author} {\bibfnamefont {R.~M.}\ \bibnamefont
  {Fernandes}}, \bibinfo {author} {\bibfnamefont {I.}~\bibnamefont {Paul}},
  \bibinfo {author} {\bibfnamefont {L.}~\bibnamefont {Chauvi{\`e}re}}, \bibinfo
  {author} {\bibfnamefont {Y.~X.}\ \bibnamefont {Yang}}, \bibinfo {author}
  {\bibfnamefont {M.~A.}\ \bibnamefont {M{\'e}asson}}, \bibinfo {author}
  {\bibfnamefont {M.}~\bibnamefont {Cazayous}}, \bibinfo {author}
  {\bibfnamefont {A.}~\bibnamefont {Sacuto}}, \bibinfo {author} {\bibfnamefont
  {D.}~\bibnamefont {Colson}}, \ and\ \bibinfo {author} {\bibfnamefont
  {A.}~\bibnamefont {Forget}},\ }\href@noop {} {\bibfield  {journal} {\bibinfo
  {journal} {Phys. Rev. Lett.}\ }\textbf {\bibinfo {volume} {111}},\ \bibinfo
  {pages} {267001} (\bibinfo {year} {2013})}\BibitemShut {NoStop}%
\bibitem [{\citenamefont {Chuang}\ \emph {et~al.}(2010)\citenamefont {Chuang},
  \citenamefont {Allan}, \citenamefont {Lee}, \citenamefont {Xie},
  \citenamefont {Ni}, \citenamefont {Bud'ko}, \citenamefont {Boebinger},
  \citenamefont {Canfield},\ and\ \citenamefont {Davis}}]{sciencedavis10}%
  \BibitemOpen
  \bibfield  {author} {\bibinfo {author} {\bibfnamefont {T.-M.}\ \bibnamefont
  {Chuang}}, \bibinfo {author} {\bibfnamefont {M.~P.}\ \bibnamefont {Allan}},
  \bibinfo {author} {\bibfnamefont {J.}~\bibnamefont {Lee}}, \bibinfo {author}
  {\bibfnamefont {Y.}~\bibnamefont {Xie}}, \bibinfo {author} {\bibfnamefont
  {N.}~\bibnamefont {Ni}}, \bibinfo {author} {\bibfnamefont {S.~L.}\
  \bibnamefont {Bud'ko}}, \bibinfo {author} {\bibfnamefont {G.~S.}\
  \bibnamefont {Boebinger}}, \bibinfo {author} {\bibfnamefont {P.~C.}\
  \bibnamefont {Canfield}}, \ and\ \bibinfo {author} {\bibfnamefont {J.~C.}\
  \bibnamefont {Davis}},\ }\href {\doibase 10.1126/science.1181083} {\bibfield
  {journal} {\bibinfo  {journal} {Science}\ }\textbf {\bibinfo {volume}
  {327}},\ \bibinfo {pages} {181} (\bibinfo {year} {2010})}\BibitemShut
  {NoStop}%
\bibitem [{\citenamefont {Allan}\ \emph {et~al.}(2013)\citenamefont {Allan},
  \citenamefont {Chuang}, \citenamefont {Massee}, \citenamefont {Xie},
  \citenamefont {Ni}, \citenamefont {Bud/'ko}, \citenamefont {Boebinger},
  \citenamefont {Wang}, \citenamefont {Dessau}, \citenamefont {Canfield},
  \citenamefont {Golden},\ and\ \citenamefont {Davis}}]{davisnatphys2013}%
  \BibitemOpen
  \bibfield  {author} {\bibinfo {author} {\bibfnamefont {M.~P.}\ \bibnamefont
  {Allan}}, \bibinfo {author} {\bibfnamefont {T.-M.}\ \bibnamefont {Chuang}},
  \bibinfo {author} {\bibfnamefont {F.}~\bibnamefont {Massee}}, \bibinfo
  {author} {\bibfnamefont {Y.}~\bibnamefont {Xie}}, \bibinfo {author}
  {\bibfnamefont {N.}~\bibnamefont {Ni}}, \bibinfo {author} {\bibfnamefont
  {S.~L.}\ \bibnamefont {Bud/'ko}}, \bibinfo {author} {\bibfnamefont {G.~S.}\
  \bibnamefont {Boebinger}}, \bibinfo {author} {\bibfnamefont {Q.}~\bibnamefont
  {Wang}}, \bibinfo {author} {\bibfnamefont {D.~S.}\ \bibnamefont {Dessau}},
  \bibinfo {author} {\bibfnamefont {P.~C.}\ \bibnamefont {Canfield}}, \bibinfo
  {author} {\bibfnamefont {M.~S.}\ \bibnamefont {Golden}}, \ and\ \bibinfo
  {author} {\bibfnamefont {J.~C.}\ \bibnamefont {Davis}},\ }\href@noop {}
  {\bibfield  {journal} {\bibinfo  {journal} {Nat. Phys.}\ }\textbf {\bibinfo
  {volume} {9}},\ \bibinfo {pages} {220} (\bibinfo {year} {2013})}\BibitemShut
  {NoStop}%
\bibitem [{\citenamefont {Rosenthal}\ \emph {et~al.}(2014)\citenamefont
  {Rosenthal}, \citenamefont {Andrade}, \citenamefont {Arguello}, \citenamefont
  {Fernandes}, \citenamefont {Xing}, \citenamefont {Wang}, \citenamefont {Jin},
  \citenamefont {Millis},\ and\ \citenamefont
  {Pasupathy}}]{rosenthalnatphys14}%
  \BibitemOpen
  \bibfield  {author} {\bibinfo {author} {\bibfnamefont {E.~P.}\ \bibnamefont
  {Rosenthal}}, \bibinfo {author} {\bibfnamefont {E.~F.}\ \bibnamefont
  {Andrade}}, \bibinfo {author} {\bibfnamefont {C.~J.}\ \bibnamefont
  {Arguello}}, \bibinfo {author} {\bibfnamefont {R.~M.}\ \bibnamefont
  {Fernandes}}, \bibinfo {author} {\bibfnamefont {L.~Y.}\ \bibnamefont {Xing}},
  \bibinfo {author} {\bibfnamefont {X.~C.}\ \bibnamefont {Wang}}, \bibinfo
  {author} {\bibfnamefont {C.~Q.}\ \bibnamefont {Jin}}, \bibinfo {author}
  {\bibfnamefont {A.~J.}\ \bibnamefont {Millis}}, \ and\ \bibinfo {author}
  {\bibfnamefont {A.~N.}\ \bibnamefont {Pasupathy}},\ }\href@noop {} {\bibfield
   {journal} {\bibinfo  {journal} {Nat Phys}\ }\textbf {\bibinfo {volume}
  {10}},\ \bibinfo {pages} {225} (\bibinfo {year} {2014})}\BibitemShut
  {NoStop}%
\bibitem [{\citenamefont {Kasahara}\ \emph {et~al.}(2012)\citenamefont
  {Kasahara}, \citenamefont {Shi}, \citenamefont {Hashimoto}, \citenamefont
  {Tonegawa}, \citenamefont {Mizukami}, \citenamefont {Shibauchi},
  \citenamefont {Sugimoto}, \citenamefont {Fukuda}, \citenamefont {Terashima},
  \citenamefont {Nevidomskyy},\ and\ \citenamefont {Matsuda}}]{matsudanat12}%
  \BibitemOpen
  \bibfield  {author} {\bibinfo {author} {\bibfnamefont {S.}~\bibnamefont
  {Kasahara}}, \bibinfo {author} {\bibfnamefont {H.~J.}\ \bibnamefont {Shi}},
  \bibinfo {author} {\bibfnamefont {K.}~\bibnamefont {Hashimoto}}, \bibinfo
  {author} {\bibfnamefont {S.}~\bibnamefont {Tonegawa}}, \bibinfo {author}
  {\bibfnamefont {Y.}~\bibnamefont {Mizukami}}, \bibinfo {author}
  {\bibfnamefont {T.}~\bibnamefont {Shibauchi}}, \bibinfo {author}
  {\bibfnamefont {K.}~\bibnamefont {Sugimoto}}, \bibinfo {author}
  {\bibfnamefont {T.}~\bibnamefont {Fukuda}}, \bibinfo {author} {\bibfnamefont
  {T.}~\bibnamefont {Terashima}}, \bibinfo {author} {\bibfnamefont {A.~H.}\
  \bibnamefont {Nevidomskyy}}, \ and\ \bibinfo {author} {\bibfnamefont
  {Y.}~\bibnamefont {Matsuda}},\ }\href@noop {} {\bibfield  {journal} {\bibinfo
   {journal} {Nature}\ }\textbf {\bibinfo {volume} {486}},\ \bibinfo {pages}
  {382} (\bibinfo {year} {2012})}\BibitemShut {NoStop}%
\bibitem [{\citenamefont {Harriger}\ \emph {et~al.}(2011)\citenamefont
  {Harriger}, \citenamefont {Luo}, \citenamefont {Liu}, \citenamefont {Frost},
  \citenamefont {Hu}, \citenamefont {Norman},\ and\ \citenamefont
  {Dai}}]{pengchengdaiprb11}%
  \BibitemOpen
  \bibfield  {author} {\bibinfo {author} {\bibfnamefont {L.~W.}\ \bibnamefont
  {Harriger}}, \bibinfo {author} {\bibfnamefont {H.~Q.}\ \bibnamefont {Luo}},
  \bibinfo {author} {\bibfnamefont {M.~S.}\ \bibnamefont {Liu}}, \bibinfo
  {author} {\bibfnamefont {C.}~\bibnamefont {Frost}}, \bibinfo {author}
  {\bibfnamefont {J.~P.}\ \bibnamefont {Hu}}, \bibinfo {author} {\bibfnamefont
  {M.~R.}\ \bibnamefont {Norman}}, \ and\ \bibinfo {author} {\bibfnamefont
  {P.}~\bibnamefont {Dai}},\ }\href@noop {} {\bibfield  {journal} {\bibinfo
  {journal} {Phys. Rev. B}\ }\textbf {\bibinfo {volume} {84}},\ \bibinfo
  {pages} {054544} (\bibinfo {year} {2011})}\BibitemShut {NoStop}%
\bibitem [{\citenamefont {Dhital}\ \emph {et~al.}(2012)\citenamefont {Dhital},
  \citenamefont {Yamani}, \citenamefont {Tian}, \citenamefont {Zeretsky},
  \citenamefont {Sefat}, \citenamefont {Wang}, \citenamefont {Birgeneau},\ and\
  \citenamefont {Wilson}}]{dhitalprl12}%
  \BibitemOpen
  \bibfield  {author} {\bibinfo {author} {\bibfnamefont {C.}~\bibnamefont
  {Dhital}}, \bibinfo {author} {\bibfnamefont {Z.}~\bibnamefont {Yamani}},
  \bibinfo {author} {\bibfnamefont {W.}~\bibnamefont {Tian}}, \bibinfo {author}
  {\bibfnamefont {J.}~\bibnamefont {Zeretsky}}, \bibinfo {author}
  {\bibfnamefont {A.~S.}\ \bibnamefont {Sefat}}, \bibinfo {author}
  {\bibfnamefont {Z.}~\bibnamefont {Wang}}, \bibinfo {author} {\bibfnamefont
  {R.~J.}\ \bibnamefont {Birgeneau}}, \ and\ \bibinfo {author} {\bibfnamefont
  {S.~D.}\ \bibnamefont {Wilson}},\ }\href@noop {} {\bibfield  {journal}
  {\bibinfo  {journal} {Phys. Rev. Lett.}\ }\textbf {\bibinfo {volume} {108}},\
  \bibinfo {pages} {087001} (\bibinfo {year} {2012})}\BibitemShut {NoStop}%
\bibitem [{\citenamefont {Song}\ \emph
  {et~al.}(2013{\natexlab{a}})\citenamefont {Song}, \citenamefont {Carr},
  \citenamefont {Lu}, \citenamefont {Zhang}, \citenamefont {Sims},
  \citenamefont {Luttrell}, \citenamefont {Chi}, \citenamefont {Zhao},
  \citenamefont {Lynn},\ and\ \citenamefont {Dai}}]{pengchengdaiprb13}%
  \BibitemOpen
  \bibfield  {author} {\bibinfo {author} {\bibfnamefont {Y.}~\bibnamefont
  {Song}}, \bibinfo {author} {\bibfnamefont {S.~V.}\ \bibnamefont {Carr}},
  \bibinfo {author} {\bibfnamefont {X.}~\bibnamefont {Lu}}, \bibinfo {author}
  {\bibfnamefont {C.}~\bibnamefont {Zhang}}, \bibinfo {author} {\bibfnamefont
  {Z.~C.}\ \bibnamefont {Sims}}, \bibinfo {author} {\bibfnamefont {N.~F.}\
  \bibnamefont {Luttrell}}, \bibinfo {author} {\bibfnamefont {S.}~\bibnamefont
  {Chi}}, \bibinfo {author} {\bibfnamefont {Y.}~\bibnamefont {Zhao}}, \bibinfo
  {author} {\bibfnamefont {J.~W.}\ \bibnamefont {Lynn}}, \ and\ \bibinfo
  {author} {\bibfnamefont {P.}~\bibnamefont {Dai}},\ }\href@noop {} {\bibfield
  {journal} {\bibinfo  {journal} {Phys. Rev. B}\ }\textbf {\bibinfo {volume}
  {87}},\ \bibinfo {pages} {184511} (\bibinfo {year}
  {2013}{\natexlab{a}})}\BibitemShut {NoStop}%
\bibitem [{\citenamefont {Ibuka}\ \emph {et~al.}(2014)\citenamefont {Ibuka},
  \citenamefont {Nambu}, \citenamefont {Yamazaki}, \citenamefont {Lumsden},\
  and\ \citenamefont {Sato}}]{ibukaphysC14}%
  \BibitemOpen
  \bibfield  {author} {\bibinfo {author} {\bibfnamefont {S.}~\bibnamefont
  {Ibuka}}, \bibinfo {author} {\bibfnamefont {Y.}~\bibnamefont {Nambu}},
  \bibinfo {author} {\bibfnamefont {T.}~\bibnamefont {Yamazaki}}, \bibinfo
  {author} {\bibfnamefont {M.~D.}\ \bibnamefont {Lumsden}}, \ and\ \bibinfo
  {author} {\bibfnamefont {T.~J.}\ \bibnamefont {Sato}},\ }\href@noop {}
  {\bibfield  {journal} {\bibinfo  {journal} {Physica C: Superconductivity}\
  }\textbf {\bibinfo {volume} {507}},\ \bibinfo {pages} {25} (\bibinfo {year}
  {2014})}\BibitemShut {NoStop}%
\bibitem [{\citenamefont {Lu}\ \emph {et~al.}(2014)\citenamefont {Lu},
  \citenamefont {Park}, \citenamefont {Zhang}, \citenamefont {Luo},
  \citenamefont {Nevidomskyy}, \citenamefont {Si},\ and\ \citenamefont
  {Dai}}]{luscience14}%
  \BibitemOpen
  \bibfield  {author} {\bibinfo {author} {\bibfnamefont {X.}~\bibnamefont
  {Lu}}, \bibinfo {author} {\bibfnamefont {J.~T.}\ \bibnamefont {Park}},
  \bibinfo {author} {\bibfnamefont {R.}~\bibnamefont {Zhang}}, \bibinfo
  {author} {\bibfnamefont {H.}~\bibnamefont {Luo}}, \bibinfo {author}
  {\bibfnamefont {A.~H.}\ \bibnamefont {Nevidomskyy}}, \bibinfo {author}
  {\bibfnamefont {Q.}~\bibnamefont {Si}}, \ and\ \bibinfo {author}
  {\bibfnamefont {P.}~\bibnamefont {Dai}},\ }\href@noop {} {\bibfield
  {journal} {\bibinfo  {journal} {Science}\ }\textbf {\bibinfo {volume}
  {345}},\ \bibinfo {pages} {657} (\bibinfo {year} {2014})}\BibitemShut
  {NoStop}%
\bibitem [{\citenamefont {Yi}\ \emph {et~al.}(2011)\citenamefont {Yi},
  \citenamefont {Lu}, \citenamefont {Chu}, \citenamefont {Analytis},
  \citenamefont {Sorini}, \citenamefont {Kemper}, \citenamefont {Moritz},
  \citenamefont {Mo}, \citenamefont {Moore}, \citenamefont {Hashimoto},
  \citenamefont {Lee}, \citenamefont {Hussain}, \citenamefont {Devereaux},
  \citenamefont {Fisher},\ and\ \citenamefont {Shen}}]{shenpnas11}%
  \BibitemOpen
  \bibfield  {author} {\bibinfo {author} {\bibfnamefont {M.}~\bibnamefont
  {Yi}}, \bibinfo {author} {\bibfnamefont {D.}~\bibnamefont {Lu}}, \bibinfo
  {author} {\bibfnamefont {J.-H.}\ \bibnamefont {Chu}}, \bibinfo {author}
  {\bibfnamefont {J.~G.}\ \bibnamefont {Analytis}}, \bibinfo {author}
  {\bibfnamefont {A.~P.}\ \bibnamefont {Sorini}}, \bibinfo {author}
  {\bibfnamefont {A.~F.}\ \bibnamefont {Kemper}}, \bibinfo {author}
  {\bibfnamefont {B.}~\bibnamefont {Moritz}}, \bibinfo {author} {\bibfnamefont
  {S.-K.}\ \bibnamefont {Mo}}, \bibinfo {author} {\bibfnamefont {R.~G.}\
  \bibnamefont {Moore}}, \bibinfo {author} {\bibfnamefont {M.}~\bibnamefont
  {Hashimoto}}, \bibinfo {author} {\bibfnamefont {W.-S.}\ \bibnamefont {Lee}},
  \bibinfo {author} {\bibfnamefont {Z.}~\bibnamefont {Hussain}}, \bibinfo
  {author} {\bibfnamefont {T.~P.}\ \bibnamefont {Devereaux}}, \bibinfo {author}
  {\bibfnamefont {I.~R.}\ \bibnamefont {Fisher}}, \ and\ \bibinfo {author}
  {\bibfnamefont {Z.-X.}\ \bibnamefont {Shen}},\ }\href@noop {} {\bibfield
  {journal} {\bibinfo  {journal} {PNAS}\ }\textbf {\bibinfo {volume} {108}},\
  \bibinfo {pages} {6878} (\bibinfo {year} {2011})}\BibitemShut {NoStop}%
\bibitem [{\citenamefont {Yi}\ \emph {et~al.}(2012)\citenamefont {Yi},
  \citenamefont {Lu}, \citenamefont {Moore}, \citenamefont {Kihou},
  \citenamefont {Lee}, \citenamefont {Iyo}, \citenamefont {Eisaki},
  \citenamefont {Yoshida}, \citenamefont {Fujimori},\ and\ \citenamefont
  {Shen}}]{shennjp12}%
  \BibitemOpen
  \bibfield  {author} {\bibinfo {author} {\bibfnamefont {M.}~\bibnamefont
  {Yi}}, \bibinfo {author} {\bibfnamefont {D.~H.}\ \bibnamefont {Lu}}, \bibinfo
  {author} {\bibfnamefont {R.~G.}\ \bibnamefont {Moore}}, \bibinfo {author}
  {\bibfnamefont {K.}~\bibnamefont {Kihou}}, \bibinfo {author} {\bibfnamefont
  {C.-H.}\ \bibnamefont {Lee}}, \bibinfo {author} {\bibfnamefont
  {A.}~\bibnamefont {Iyo}}, \bibinfo {author} {\bibfnamefont {H.}~\bibnamefont
  {Eisaki}}, \bibinfo {author} {\bibfnamefont {T.}~\bibnamefont {Yoshida}},
  \bibinfo {author} {\bibfnamefont {A.}~\bibnamefont {Fujimori}}, \ and\
  \bibinfo {author} {\bibfnamefont {Z.-X.}\ \bibnamefont {Shen}},\ }\href@noop
  {} {\bibfield  {journal} {\bibinfo  {journal} {New J. Phys.}\ }\textbf
  {\bibinfo {volume} {14}},\ \bibinfo {pages} {073019} (\bibinfo {year}
  {2012})}\BibitemShut {NoStop}%
\bibitem [{\citenamefont {Kim}\ \emph {et~al.}(2013)\citenamefont {Kim},
  \citenamefont {Jung}, \citenamefont {Han}, \citenamefont {Choi},
  \citenamefont {Chen}, \citenamefont {Devereaux}, \citenamefont {Chainani},
  \citenamefont {Miyawaki}, \citenamefont {Takata}, \citenamefont {Tanaka},
  \citenamefont {Oura}, \citenamefont {Shin}, \citenamefont {Singh},
  \citenamefont {Lee}, \citenamefont {Kim},\ and\ \citenamefont {Kim}}]{kim13}%
  \BibitemOpen
  \bibfield  {author} {\bibinfo {author} {\bibfnamefont {Y.~K.}\ \bibnamefont
  {Kim}}, \bibinfo {author} {\bibfnamefont {W.~S.}\ \bibnamefont {Jung}},
  \bibinfo {author} {\bibfnamefont {G.~R.}\ \bibnamefont {Han}}, \bibinfo
  {author} {\bibfnamefont {K.-Y.}\ \bibnamefont {Choi}}, \bibinfo {author}
  {\bibfnamefont {C.-C.}\ \bibnamefont {Chen}}, \bibinfo {author}
  {\bibfnamefont {T.~P.}\ \bibnamefont {Devereaux}}, \bibinfo {author}
  {\bibfnamefont {A.}~\bibnamefont {Chainani}}, \bibinfo {author}
  {\bibfnamefont {J.}~\bibnamefont {Miyawaki}}, \bibinfo {author}
  {\bibfnamefont {Y.}~\bibnamefont {Takata}}, \bibinfo {author} {\bibfnamefont
  {Y.}~\bibnamefont {Tanaka}}, \bibinfo {author} {\bibfnamefont
  {M.}~\bibnamefont {Oura}}, \bibinfo {author} {\bibfnamefont {S.}~\bibnamefont
  {Shin}}, \bibinfo {author} {\bibfnamefont {A.~P.}\ \bibnamefont {Singh}},
  \bibinfo {author} {\bibfnamefont {H.~G.}\ \bibnamefont {Lee}}, \bibinfo
  {author} {\bibfnamefont {J.-Y.}\ \bibnamefont {Kim}}, \ and\ \bibinfo
  {author} {\bibfnamefont {C.}~\bibnamefont {Kim}},\ }\href@noop {} {\bibfield
  {journal} {\bibinfo  {journal} {Phys. Rev. Lett.}\ }\textbf {\bibinfo
  {volume} {111}},\ \bibinfo {pages} {217001} (\bibinfo {year}
  {2013})}\BibitemShut {NoStop}%
\bibitem [{\citenamefont {Ishida}\ \emph {et~al.}(2013)\citenamefont {Ishida},
  \citenamefont {Nakajima}, \citenamefont {Liang}, \citenamefont {Kihou},
  \citenamefont {Lee}, \citenamefont {Iyo}, \citenamefont {Eisaki},
  \citenamefont {Kakeshita}, \citenamefont {Tomioka}, \citenamefont {Ito},\
  and\ \citenamefont {Uchida}}]{uchidaprl13}%
  \BibitemOpen
  \bibfield  {author} {\bibinfo {author} {\bibfnamefont {S.}~\bibnamefont
  {Ishida}}, \bibinfo {author} {\bibfnamefont {M.}~\bibnamefont {Nakajima}},
  \bibinfo {author} {\bibfnamefont {T.}~\bibnamefont {Liang}}, \bibinfo
  {author} {\bibfnamefont {K.}~\bibnamefont {Kihou}}, \bibinfo {author}
  {\bibfnamefont {C.~H.}\ \bibnamefont {Lee}}, \bibinfo {author} {\bibfnamefont
  {A.}~\bibnamefont {Iyo}}, \bibinfo {author} {\bibfnamefont {H.}~\bibnamefont
  {Eisaki}}, \bibinfo {author} {\bibfnamefont {T.}~\bibnamefont {Kakeshita}},
  \bibinfo {author} {\bibfnamefont {Y.}~\bibnamefont {Tomioka}}, \bibinfo
  {author} {\bibfnamefont {T.}~\bibnamefont {Ito}}, \ and\ \bibinfo {author}
  {\bibfnamefont {S.}~\bibnamefont {Uchida}},\ }\href@noop {} {\bibfield
  {journal} {\bibinfo  {journal} {Phys. Rev. Lett.}\ }\textbf {\bibinfo
  {volume} {110}},\ \bibinfo {pages} {207001} (\bibinfo {year}
  {2013})}\BibitemShut {NoStop}%
\bibitem [{\citenamefont {Gastiasoro}\ \emph {et~al.}(2014)\citenamefont
  {Gastiasoro}, \citenamefont {Paul}, \citenamefont {Wang}, \citenamefont
  {Hirschfeld},\ and\ \citenamefont {Andersen}}]{andersenprl13}%
  \BibitemOpen
  \bibfield  {author} {\bibinfo {author} {\bibfnamefont {M.~N.}\ \bibnamefont
  {Gastiasoro}}, \bibinfo {author} {\bibfnamefont {I.}~\bibnamefont {Paul}},
  \bibinfo {author} {\bibfnamefont {Y.}~\bibnamefont {Wang}}, \bibinfo {author}
  {\bibfnamefont {P.~J.}\ \bibnamefont {Hirschfeld}}, \ and\ \bibinfo {author}
  {\bibfnamefont {B.~M.}\ \bibnamefont {Andersen}},\ }\href@noop {} {\bibfield
  {journal} {\bibinfo  {journal} {Phys. Rev. Lett.}\ }\textbf {\bibinfo
  {volume} {113}},\ \bibinfo {pages} {127001} (\bibinfo {year}
  {2014})}\BibitemShut {NoStop}%
\bibitem [{\citenamefont {Wang}\ \emph {et~al.}(2015)\citenamefont {Wang},
  \citenamefont {Gastiasoro}, \citenamefont {Andersen}, \citenamefont
  {Tomi\ifmmode~\acute{c}\else \'{c}\fi{}}, \citenamefont {Jeschke},
  \citenamefont {Valent\'{i}}, \citenamefont {Paul},\ and\ \citenamefont
  {Hirschfeld}}]{hirschfeldarXiv14}%
  \BibitemOpen
  \bibfield  {author} {\bibinfo {author} {\bibfnamefont {Y.}~\bibnamefont
  {Wang}}, \bibinfo {author} {\bibfnamefont {M.~N.}\ \bibnamefont
  {Gastiasoro}}, \bibinfo {author} {\bibfnamefont {B.~M.}\ \bibnamefont
  {Andersen}}, \bibinfo {author} {\bibfnamefont {M.}~\bibnamefont
  {Tomi\ifmmode~\acute{c}\else \'{c}\fi{}}}, \bibinfo {author} {\bibfnamefont
  {H.~O.}\ \bibnamefont {Jeschke}}, \bibinfo {author} {\bibfnamefont
  {R.}~\bibnamefont {Valent\'{i}}}, \bibinfo {author} {\bibfnamefont
  {I.}~\bibnamefont {Paul}}, \ and\ \bibinfo {author} {\bibfnamefont {P.~J.}\
  \bibnamefont {Hirschfeld}},\ }\href {\doibase 10.1103/PhysRevLett.114.097003}
  {\bibfield  {journal} {\bibinfo  {journal} {Phys. Rev. Lett.}\ }\textbf
  {\bibinfo {volume} {114}},\ \bibinfo {pages} {097003} (\bibinfo {year}
  {2015})}\BibitemShut {NoStop}%
\bibitem [{\citenamefont {Fang}\ \emph {et~al.}(2008)\citenamefont {Fang},
  \citenamefont {Yao}, \citenamefont {Tsai}, \citenamefont {Hu},\ and\
  \citenamefont {Kivelson}}]{kivelson08}%
  \BibitemOpen
  \bibfield  {author} {\bibinfo {author} {\bibfnamefont {C.}~\bibnamefont
  {Fang}}, \bibinfo {author} {\bibfnamefont {H.}~\bibnamefont {Yao}}, \bibinfo
  {author} {\bibfnamefont {W.-F.}\ \bibnamefont {Tsai}}, \bibinfo {author}
  {\bibfnamefont {J.}~\bibnamefont {Hu}}, \ and\ \bibinfo {author}
  {\bibfnamefont {S.~A.}\ \bibnamefont {Kivelson}},\ }\href {\doibase
  10.1103/PhysRevB.77.224509} {\bibfield  {journal} {\bibinfo  {journal} {Phys.
  Rev. B}\ }\textbf {\bibinfo {volume} {77}},\ \bibinfo {pages} {224509}
  (\bibinfo {year} {2008})}\BibitemShut {NoStop}%
\bibitem [{\citenamefont {Xu}\ \emph {et~al.}(2008)\citenamefont {Xu},
  \citenamefont {M{\"u}ller},\ and\ \citenamefont {Sachdev}}]{sachdevprb08}%
  \BibitemOpen
  \bibfield  {author} {\bibinfo {author} {\bibfnamefont {C.}~\bibnamefont
  {Xu}}, \bibinfo {author} {\bibfnamefont {M.}~\bibnamefont {M{\"u}ller}}, \
  and\ \bibinfo {author} {\bibfnamefont {S.}~\bibnamefont {Sachdev}},\
  }\href@noop {} {\bibfield  {journal} {\bibinfo  {journal} {Phys. Rev. B}\
  }\textbf {\bibinfo {volume} {78}},\ \bibinfo {pages} {020501} (\bibinfo
  {year} {2008})}\BibitemShut {NoStop}%
\bibitem [{\citenamefont {Lee}\ \emph {et~al.}(2009)\citenamefont {Lee},
  \citenamefont {Yin},\ and\ \citenamefont {Ku}}]{leeyinku09}%
  \BibitemOpen
  \bibfield  {author} {\bibinfo {author} {\bibfnamefont {C.-C.}\ \bibnamefont
  {Lee}}, \bibinfo {author} {\bibfnamefont {W.-G.}\ \bibnamefont {Yin}}, \ and\
  \bibinfo {author} {\bibfnamefont {W.}~\bibnamefont {Ku}},\ }\href {\doibase
  10.1103/PhysRevLett.103.267001} {\bibfield  {journal} {\bibinfo  {journal}
  {Phys. Rev. Lett.}\ }\textbf {\bibinfo {volume} {103}},\ \bibinfo {pages}
  {267001} (\bibinfo {year} {2009})}\BibitemShut {NoStop}%
\bibitem [{\citenamefont {Valenzuela}\ \emph {et~al.}(2010)\citenamefont
  {Valenzuela}, \citenamefont {Bascones},\ and\ \citenamefont
  {Calder\'on}}]{nosotrasprl10-2}%
  \BibitemOpen
  \bibfield  {author} {\bibinfo {author} {\bibfnamefont {B.}~\bibnamefont
  {Valenzuela}}, \bibinfo {author} {\bibfnamefont {E.}~\bibnamefont
  {Bascones}}, \ and\ \bibinfo {author} {\bibfnamefont {M.~J.}\ \bibnamefont
  {Calder\'on}},\ }\href@noop {} {\bibfield  {journal} {\bibinfo  {journal}
  {Phys. Rev. Lett.}\ }\textbf {\bibinfo {volume} {105}},\ \bibinfo {pages}
  {207202} (\bibinfo {year} {2010})}\BibitemShut {NoStop}%
\bibitem [{\citenamefont {Yin}\ and\ \citenamefont
  {Pickett}(2010)}]{yinpickettprb10}%
  \BibitemOpen
  \bibfield  {author} {\bibinfo {author} {\bibfnamefont {Z.~P.}\ \bibnamefont
  {Yin}}\ and\ \bibinfo {author} {\bibfnamefont {W.~E.}\ \bibnamefont
  {Pickett}},\ }\href@noop {} {\bibfield  {journal} {\bibinfo  {journal}
  {Physical Review B}\ }\textbf {\bibinfo {volume} {81}},\ \bibinfo {pages}
  {174534} (\bibinfo {year} {2010})}\BibitemShut {NoStop}%
\bibitem [{\citenamefont {Lv}\ \emph {et~al.}(2010)\citenamefont {Lv},
  \citenamefont {Kr{\"u}ger},\ and\ \citenamefont
  {Phillips}}]{lvphillipsprb10}%
  \BibitemOpen
  \bibfield  {author} {\bibinfo {author} {\bibfnamefont {W.}~\bibnamefont
  {Lv}}, \bibinfo {author} {\bibfnamefont {F.}~\bibnamefont {Kr{\"u}ger}}, \
  and\ \bibinfo {author} {\bibfnamefont {P.}~\bibnamefont {Phillips}},\
  }\href@noop {} {\bibfield  {journal} {\bibinfo  {journal} {Phys. Rev. B}\
  }\textbf {\bibinfo {volume} {82}},\ \bibinfo {pages} {045125} (\bibinfo
  {year} {2010})}\BibitemShut {NoStop}%
\bibitem [{\citenamefont {Chen}\ \emph {et~al.}(2010)\citenamefont {Chen},
  \citenamefont {Maciejko}, \citenamefont {Sorini}, \citenamefont {Moritz},
  \citenamefont {Singh},\ and\ \citenamefont {Devereaux}}]{chen_devereaux10}%
  \BibitemOpen
  \bibfield  {author} {\bibinfo {author} {\bibfnamefont {C.-C.}\ \bibnamefont
  {Chen}}, \bibinfo {author} {\bibfnamefont {J.}~\bibnamefont {Maciejko}},
  \bibinfo {author} {\bibfnamefont {A.~P.}\ \bibnamefont {Sorini}}, \bibinfo
  {author} {\bibfnamefont {B.}~\bibnamefont {Moritz}}, \bibinfo {author}
  {\bibfnamefont {R.~R.~P.}\ \bibnamefont {Singh}}, \ and\ \bibinfo {author}
  {\bibfnamefont {T.~P.}\ \bibnamefont {Devereaux}},\ }\href {\doibase
  10.1103/PhysRevB.82.100504} {\bibfield  {journal} {\bibinfo  {journal} {Phys.
  Rev. B}\ }\textbf {\bibinfo {volume} {82}},\ \bibinfo {pages} {100504}
  (\bibinfo {year} {2010})}\BibitemShut {NoStop}%
\bibitem [{\citenamefont {Laad}\ and\ \citenamefont
  {Craco}(2011)}]{laadcracoprb11}%
  \BibitemOpen
  \bibfield  {author} {\bibinfo {author} {\bibfnamefont {M.~S.}\ \bibnamefont
  {Laad}}\ and\ \bibinfo {author} {\bibfnamefont {L.}~\bibnamefont {Craco}},\
  }\href@noop {} {\bibfield  {journal} {\bibinfo  {journal} {Phys. Rev. B}\
  }\textbf {\bibinfo {volume} {84}},\ \bibinfo {pages} {054530} (\bibinfo
  {year} {2011})}\BibitemShut {NoStop}%
\bibitem [{\citenamefont {Cano}\ \emph {et~al.}(2010)\citenamefont {Cano},
  \citenamefont {Civelli}, \citenamefont {Eremin},\ and\ \citenamefont
  {Paul}}]{canoprb10}%
  \BibitemOpen
  \bibfield  {author} {\bibinfo {author} {\bibfnamefont {A.}~\bibnamefont
  {Cano}}, \bibinfo {author} {\bibfnamefont {M.}~\bibnamefont {Civelli}},
  \bibinfo {author} {\bibfnamefont {I.}~\bibnamefont {Eremin}}, \ and\ \bibinfo
  {author} {\bibfnamefont {I.}~\bibnamefont {Paul}},\ }\href@noop {} {\bibfield
   {journal} {\bibinfo  {journal} {Phys. Rev. B}\ }\textbf {\bibinfo {volume}
  {82}},\ \bibinfo {pages} {020408} (\bibinfo {year} {2010})}\BibitemShut
  {NoStop}%
\bibitem [{\citenamefont {Cano}(2011)}]{canoprb11}%
  \BibitemOpen
  \bibfield  {author} {\bibinfo {author} {\bibfnamefont {A.}~\bibnamefont
  {Cano}},\ }\href@noop {} {\bibfield  {journal} {\bibinfo  {journal} {Phys.
  Rev. B}\ }\textbf {\bibinfo {volume} {84}},\ \bibinfo {pages} {012504}
  (\bibinfo {year} {2011})}\BibitemShut {NoStop}%
\bibitem [{\citenamefont {Cano}\ and\ \citenamefont {Paul}(2012)}]{canoprb12}%
  \BibitemOpen
  \bibfield  {author} {\bibinfo {author} {\bibfnamefont {A.}~\bibnamefont
  {Cano}}\ and\ \bibinfo {author} {\bibfnamefont {I.}~\bibnamefont {Paul}},\
  }\href@noop {} {\bibfield  {journal} {\bibinfo  {journal} {Phys. Rev. B}\
  }\textbf {\bibinfo {volume} {85}},\ \bibinfo {pages} {155133} (\bibinfo
  {year} {2012})}\BibitemShut {NoStop}%
\bibitem [{\citenamefont {Fernandes}\ \emph {et~al.}(2012)\citenamefont
  {Fernandes}, \citenamefont {Chubukov}, \citenamefont {Knolle}, \citenamefont
  {Eremin},\ and\ \citenamefont {Schmalian}}]{schmalianprb12}%
  \BibitemOpen
  \bibfield  {author} {\bibinfo {author} {\bibfnamefont {R.~M.}\ \bibnamefont
  {Fernandes}}, \bibinfo {author} {\bibfnamefont {A.~V.}\ \bibnamefont
  {Chubukov}}, \bibinfo {author} {\bibfnamefont {J.}~\bibnamefont {Knolle}},
  \bibinfo {author} {\bibfnamefont {I.}~\bibnamefont {Eremin}}, \ and\ \bibinfo
  {author} {\bibfnamefont {J.}~\bibnamefont {Schmalian}},\ }\href {\doibase
  10.1103/PhysRevB.85.024534} {\bibfield  {journal} {\bibinfo  {journal} {Phys.
  Rev. B}\ }\textbf {\bibinfo {volume} {85}},\ \bibinfo {pages} {024534}
  (\bibinfo {year} {2012})}\BibitemShut {NoStop}%
\bibitem [{\citenamefont {Fernandes}\ and\ \citenamefont
  {Schmalian}(2012)}]{Fernandesreview12}%
  \BibitemOpen
  \bibfield  {author} {\bibinfo {author} {\bibfnamefont {R.~M.}\ \bibnamefont
  {Fernandes}}\ and\ \bibinfo {author} {\bibfnamefont {J.}~\bibnamefont
  {Schmalian}},\ }\href@noop {} {\bibfield  {journal} {\bibinfo  {journal}
  {Supercond. Sci. Technol.}\ }\textbf {\bibinfo {volume} {25}},\ \bibinfo
  {pages} {084005} (\bibinfo {year} {2012})}\BibitemShut {NoStop}%
\bibitem [{\citenamefont {Liang}\ \emph {et~al.}(2013)\citenamefont {Liang},
  \citenamefont {Moreo},\ and\ \citenamefont {Dagotto}}]{liangprl13}%
  \BibitemOpen
  \bibfield  {author} {\bibinfo {author} {\bibfnamefont {S.}~\bibnamefont
  {Liang}}, \bibinfo {author} {\bibfnamefont {A.}~\bibnamefont {Moreo}}, \ and\
  \bibinfo {author} {\bibfnamefont {E.}~\bibnamefont {Dagotto}},\ }\href@noop
  {} {\bibfield  {journal} {\bibinfo  {journal} {Phys. Rev. Lett.}\ }\textbf
  {\bibinfo {volume} {111}},\ \bibinfo {pages} {047004} (\bibinfo {year}
  {2013})}\BibitemShut {NoStop}%
\bibitem [{\citenamefont {Liang}\ \emph {et~al.}(2014)\citenamefont {Liang},
  \citenamefont {Mukherjee}, \citenamefont {Patel}, \citenamefont {Dagotto},\
  and\ \citenamefont {Moreo}}]{liangarXiv14}%
  \BibitemOpen
  \bibfield  {author} {\bibinfo {author} {\bibfnamefont {S.}~\bibnamefont
  {Liang}}, \bibinfo {author} {\bibfnamefont {A.}~\bibnamefont {Mukherjee}},
  \bibinfo {author} {\bibfnamefont {N.~D.}\ \bibnamefont {Patel}}, \bibinfo
  {author} {\bibfnamefont {E.}~\bibnamefont {Dagotto}}, \ and\ \bibinfo
  {author} {\bibfnamefont {A.}~\bibnamefont {Moreo}},\ }\href@noop {}
  {}\bibinfo {howpublished} {arXiv:1405.6395} (\bibinfo {year}
  {2014})\BibitemShut {NoStop}%
\bibitem [{\citenamefont {Paul}(2014)}]{ipaulprb14}%
  \BibitemOpen
  \bibfield  {author} {\bibinfo {author} {\bibfnamefont {I.}~\bibnamefont
  {Paul}},\ }\href@noop {} {\bibfield  {journal} {\bibinfo  {journal} {Phys.
  Rev. B}\ }\textbf {\bibinfo {volume} {90}},\ \bibinfo {pages} {115102}
  (\bibinfo {year} {2014})}\BibitemShut {NoStop}%
\bibitem [{\citenamefont {Fernandes}\ \emph {et~al.}(2014)\citenamefont
  {Fernandes}, \citenamefont {Chubukov},\ and\ \citenamefont
  {Schmalian}}]{fernandesnatphys14}%
  \BibitemOpen
  \bibfield  {author} {\bibinfo {author} {\bibfnamefont {R.~M.}\ \bibnamefont
  {Fernandes}}, \bibinfo {author} {\bibfnamefont {A.~V.}\ \bibnamefont
  {Chubukov}}, \ and\ \bibinfo {author} {\bibfnamefont {J.}~\bibnamefont
  {Schmalian}},\ }\href@noop {} {\bibfield  {journal} {\bibinfo  {journal} {Nat
  Phys}\ }\textbf {\bibinfo {volume} {10}},\ \bibinfo {pages} {97} (\bibinfo
  {year} {2014})}\BibitemShut {NoStop}%
\bibitem [{\citenamefont {Davis}\ and\ \citenamefont
  {Hirschfeld}(2014)}]{hirschfeldnatphys14}%
  \BibitemOpen
  \bibfield  {author} {\bibinfo {author} {\bibfnamefont {J.~C.}\ \bibnamefont
  {Davis}}\ and\ \bibinfo {author} {\bibfnamefont {P.~J.}\ \bibnamefont
  {Hirschfeld}},\ }\href@noop {} {\bibfield  {journal} {\bibinfo  {journal}
  {Nat Phys}\ }\textbf {\bibinfo {volume} {10}},\ \bibinfo {pages} {184}
  (\bibinfo {year} {2014})}\BibitemShut {NoStop}%
\bibitem [{\citenamefont {Baek}\ \emph {et~al.}(2015)\citenamefont {Baek},
  \citenamefont {Efremov}, \citenamefont {Ok}, \citenamefont {Kim},
  \citenamefont {van~den Brink},\ and\ \citenamefont
  {B\"uchner}}]{buchner_nature15}%
  \BibitemOpen
  \bibfield  {author} {\bibinfo {author} {\bibfnamefont {S.-H.}\ \bibnamefont
  {Baek}}, \bibinfo {author} {\bibfnamefont {D.~V.}\ \bibnamefont {Efremov}},
  \bibinfo {author} {\bibfnamefont {J.~M.}\ \bibnamefont {Ok}}, \bibinfo
  {author} {\bibfnamefont {J.~S.}\ \bibnamefont {Kim}}, \bibinfo {author}
  {\bibfnamefont {J.}~\bibnamefont {van~den Brink}}, \ and\ \bibinfo {author}
  {\bibfnamefont {B.}~\bibnamefont {B\"uchner}},\ }\href@noop {} {\bibfield
  {journal} {\bibinfo  {journal} {Nature Materials}\ }\textbf {\bibinfo
  {volume} {14}},\ \bibinfo {pages} {210–214} (\bibinfo {year}
  {2015})}\BibitemShut {NoStop}%
\bibitem [{\citenamefont {Yu}\ \emph {et~al.}(2009)\citenamefont {Yu},
  \citenamefont {Trinh}, \citenamefont {Moreo}, \citenamefont {Daghofer},
  \citenamefont {Riera}, \citenamefont {Haas},\ and\ \citenamefont
  {Dagotto}}]{yu09}%
  \BibitemOpen
  \bibfield  {author} {\bibinfo {author} {\bibfnamefont {R.}~\bibnamefont
  {Yu}}, \bibinfo {author} {\bibfnamefont {K.}~\bibnamefont {Trinh}}, \bibinfo
  {author} {\bibfnamefont {A.}~\bibnamefont {Moreo}}, \bibinfo {author}
  {\bibfnamefont {M.}~\bibnamefont {Daghofer}}, \bibinfo {author}
  {\bibfnamefont {J.~A.}\ \bibnamefont {Riera}}, \bibinfo {author}
  {\bibfnamefont {S.}~\bibnamefont {Haas}}, \ and\ \bibinfo {author}
  {\bibfnamefont {E.}~\bibnamefont {Dagotto}},\ }\href@noop {} {\bibfield
  {journal} {\bibinfo  {journal} {Phys. Rev. B}\ }\textbf {\bibinfo {volume}
  {79}},\ \bibinfo {pages} {104510} (\bibinfo {year} {2009})}\BibitemShut
  {NoStop}%
\bibitem [{\citenamefont {Bascones}\ \emph {et~al.}(2010)\citenamefont
  {Bascones}, \citenamefont {Calder\'on},\ and\ \citenamefont
  {Valenzuela}}]{nosotrasprl10}%
  \BibitemOpen
  \bibfield  {author} {\bibinfo {author} {\bibfnamefont {E.}~\bibnamefont
  {Bascones}}, \bibinfo {author} {\bibfnamefont {M.~J.}\ \bibnamefont
  {Calder\'on}}, \ and\ \bibinfo {author} {\bibfnamefont {B.}~\bibnamefont
  {Valenzuela}},\ }\href@noop {} {\bibfield  {journal} {\bibinfo  {journal}
  {Phys. Rev. Lett.}\ }\textbf {\bibinfo {volume} {104}},\ \bibinfo {pages}
  {227201} (\bibinfo {year} {2010})}\BibitemShut {NoStop}%
\bibitem [{\citenamefont {Daghofer}\ \emph {et~al.}(2010)\citenamefont
  {Daghofer}, \citenamefont {Nicholson}, \citenamefont {Moreo},\ and\
  \citenamefont {Dagotto}}]{daghofer10}%
  \BibitemOpen
  \bibfield  {author} {\bibinfo {author} {\bibfnamefont {M.}~\bibnamefont
  {Daghofer}}, \bibinfo {author} {\bibfnamefont {A.}~\bibnamefont {Nicholson}},
  \bibinfo {author} {\bibfnamefont {A.}~\bibnamefont {Moreo}}, \ and\ \bibinfo
  {author} {\bibfnamefont {E.}~\bibnamefont {Dagotto}},\ }\href {\doibase
  10.1103/PhysRevB.81.014511} {\bibfield  {journal} {\bibinfo  {journal} {Phys.
  Rev. B}\ }\textbf {\bibinfo {volume} {81}},\ \bibinfo {pages} {014511}
  (\bibinfo {year} {2010})}\BibitemShut {NoStop}%
\bibitem [{\citenamefont {Yin}\ \emph {et~al.}(2011)\citenamefont {Yin},
  \citenamefont {Haule},\ and\ \citenamefont {Kotliar}}]{yin11}%
  \BibitemOpen
  \bibfield  {author} {\bibinfo {author} {\bibfnamefont {Z.~P.}\ \bibnamefont
  {Yin}}, \bibinfo {author} {\bibfnamefont {K.}~\bibnamefont {Haule}}, \ and\
  \bibinfo {author} {\bibfnamefont {G.}~\bibnamefont {Kotliar}},\ }\href@noop
  {} {\bibfield  {journal} {\bibinfo  {journal} {Nature Physics}\ }\textbf
  {\bibinfo {volume} {7}},\ \bibinfo {pages} {294} (\bibinfo {year}
  {2011})}\BibitemShut {NoStop}%
\bibitem [{\citenamefont {Graser}\ \emph {et~al.}(2009)\citenamefont {Graser},
  \citenamefont {Maier}, \citenamefont {Hirschfeld},\ and\ \citenamefont
  {Scalapino}}]{graser09}%
  \BibitemOpen
  \bibfield  {author} {\bibinfo {author} {\bibfnamefont {S.}~\bibnamefont
  {Graser}}, \bibinfo {author} {\bibfnamefont {T.}~\bibnamefont {Maier}},
  \bibinfo {author} {\bibfnamefont {P.}~\bibnamefont {Hirschfeld}}, \ and\
  \bibinfo {author} {\bibfnamefont {D.}~\bibnamefont {Scalapino}},\ }\href@noop
  {} {\bibfield  {journal} {\bibinfo  {journal} {New J. Phys.}\ }\textbf
  {\bibinfo {volume} {11}},\ \bibinfo {pages} {025016} (\bibinfo {year}
  {2009})}\BibitemShut {NoStop}%
\bibitem [{\citenamefont {Park}\ \emph {et~al.}(2010)\citenamefont {Park},
  \citenamefont {Inosov}, \citenamefont {Yaresko}, \citenamefont {Graser},
  \citenamefont {Sun}, \citenamefont {Bourges}, \citenamefont {Sidis},
  \citenamefont {Li}, \citenamefont {Kim}, \citenamefont {Haug}, \citenamefont
  {Ivanov}, \citenamefont {Hradil}, \citenamefont {Schneidewind}, \citenamefont
  {Link}, \citenamefont {Faulhaber}, \citenamefont {Glavatskyy}, \citenamefont
  {Lin}, \citenamefont {Keimer},\ and\ \citenamefont {Hinkov}}]{hinkovprb10}%
  \BibitemOpen
  \bibfield  {author} {\bibinfo {author} {\bibfnamefont {J.~T.}\ \bibnamefont
  {Park}}, \bibinfo {author} {\bibfnamefont {D.~S.}\ \bibnamefont {Inosov}},
  \bibinfo {author} {\bibfnamefont {A.}~\bibnamefont {Yaresko}}, \bibinfo
  {author} {\bibfnamefont {S.}~\bibnamefont {Graser}}, \bibinfo {author}
  {\bibfnamefont {D.~L.}\ \bibnamefont {Sun}}, \bibinfo {author} {\bibfnamefont
  {P.}~\bibnamefont {Bourges}}, \bibinfo {author} {\bibfnamefont
  {Y.}~\bibnamefont {Sidis}}, \bibinfo {author} {\bibfnamefont
  {Y.}~\bibnamefont {Li}}, \bibinfo {author} {\bibfnamefont {J.~H.}\
  \bibnamefont {Kim}}, \bibinfo {author} {\bibfnamefont {D.}~\bibnamefont
  {Haug}}, \bibinfo {author} {\bibfnamefont {A.}~\bibnamefont {Ivanov}},
  \bibinfo {author} {\bibfnamefont {K.}~\bibnamefont {Hradil}}, \bibinfo
  {author} {\bibfnamefont {A.}~\bibnamefont {Schneidewind}}, \bibinfo {author}
  {\bibfnamefont {P.}~\bibnamefont {Link}}, \bibinfo {author} {\bibfnamefont
  {E.}~\bibnamefont {Faulhaber}}, \bibinfo {author} {\bibfnamefont
  {I.}~\bibnamefont {Glavatskyy}}, \bibinfo {author} {\bibfnamefont {C.~T.}\
  \bibnamefont {Lin}}, \bibinfo {author} {\bibfnamefont {B.}~\bibnamefont
  {Keimer}}, \ and\ \bibinfo {author} {\bibfnamefont {V.}~\bibnamefont
  {Hinkov}},\ }\href@noop {} {\bibfield  {journal} {\bibinfo  {journal} {Phys.
  Rev. B}\ }\textbf {\bibinfo {volume} {82}},\ \bibinfo {pages} {134503}
  (\bibinfo {year} {2010})}\BibitemShut {NoStop}%
\bibitem [{\citenamefont {Park}\ \emph {et~al.}(2011)\citenamefont {Park},
  \citenamefont {Haule},\ and\ \citenamefont {Kotliar}}]{kotliarprl11}%
  \BibitemOpen
  \bibfield  {author} {\bibinfo {author} {\bibfnamefont {H.}~\bibnamefont
  {Park}}, \bibinfo {author} {\bibfnamefont {K.}~\bibnamefont {Haule}}, \ and\
  \bibinfo {author} {\bibfnamefont {G.}~\bibnamefont {Kotliar}},\ }\href@noop
  {} {\bibfield  {journal} {\bibinfo  {journal} {Phys. Rev. Lett.}\ }\textbf
  {\bibinfo {volume} {107}},\ \bibinfo {pages} {137007} (\bibinfo {year}
  {2011})}\BibitemShut {NoStop}%
\bibitem [{\citenamefont {Lorenzana}\ \emph {et~al.}(2008)\citenamefont
  {Lorenzana}, \citenamefont {Seibold}, \citenamefont {Ortix},\ and\
  \citenamefont {Grilli}}]{lorenzanaprl08}%
  \BibitemOpen
  \bibfield  {author} {\bibinfo {author} {\bibfnamefont {J.}~\bibnamefont
  {Lorenzana}}, \bibinfo {author} {\bibfnamefont {G.}~\bibnamefont {Seibold}},
  \bibinfo {author} {\bibfnamefont {C.}~\bibnamefont {Ortix}}, \ and\ \bibinfo
  {author} {\bibfnamefont {M.}~\bibnamefont {Grilli}},\ }\href@noop {}
  {\bibfield  {journal} {\bibinfo  {journal} {Phys. Rev. Lett.}\ }\textbf
  {\bibinfo {volume} {101}},\ \bibinfo {pages} {186402} (\bibinfo {year}
  {2008})}\BibitemShut {NoStop}%
\bibitem [{\citenamefont {Knolle}\ \emph {et~al.}(2010)\citenamefont {Knolle},
  \citenamefont {Eremin}, \citenamefont {Chubukov},\ and\ \citenamefont
  {Moessner}}]{chubukovprb10}%
  \BibitemOpen
  \bibfield  {author} {\bibinfo {author} {\bibfnamefont {J.}~\bibnamefont
  {Knolle}}, \bibinfo {author} {\bibfnamefont {I.}~\bibnamefont {Eremin}},
  \bibinfo {author} {\bibfnamefont {A.~V.}\ \bibnamefont {Chubukov}}, \ and\
  \bibinfo {author} {\bibfnamefont {R.}~\bibnamefont {Moessner}},\ }\href@noop
  {} {\bibfield  {journal} {\bibinfo  {journal} {Phys. Rev. B}\ }\textbf
  {\bibinfo {volume} {81}},\ \bibinfo {pages} {140506} (\bibinfo {year}
  {2010})}\BibitemShut {NoStop}%
\bibitem [{\citenamefont {Capati}\ \emph {et~al.}(2011)\citenamefont {Capati},
  \citenamefont {Grilli},\ and\ \citenamefont {Lorenzana}}]{lorenzanaprb11}%
  \BibitemOpen
  \bibfield  {author} {\bibinfo {author} {\bibfnamefont {M.}~\bibnamefont
  {Capati}}, \bibinfo {author} {\bibfnamefont {M.}~\bibnamefont {Grilli}}, \
  and\ \bibinfo {author} {\bibfnamefont {J.}~\bibnamefont {Lorenzana}},\
  }\href@noop {} {\bibfield  {journal} {\bibinfo  {journal} {Phys. Rev. B}\
  }\textbf {\bibinfo {volume} {84}},\ \bibinfo {pages} {214520} (\bibinfo
  {year} {2011})}\BibitemShut {NoStop}%
\bibitem [{\citenamefont {Brydon}\ \emph {et~al.}(2011)\citenamefont {Brydon},
  \citenamefont {Schmiedt},\ and\ \citenamefont {Timm}}]{brydonprb11}%
  \BibitemOpen
  \bibfield  {author} {\bibinfo {author} {\bibfnamefont {P.~M.~R.}\
  \bibnamefont {Brydon}}, \bibinfo {author} {\bibfnamefont {J.}~\bibnamefont
  {Schmiedt}}, \ and\ \bibinfo {author} {\bibfnamefont {C.}~\bibnamefont
  {Timm}},\ }\href@noop {} {\bibfield  {journal} {\bibinfo  {journal} {Phys.
  Rev. B}\ }\textbf {\bibinfo {volume} {84}},\ \bibinfo {pages} {214510}
  (\bibinfo {year} {2011})}\BibitemShut {NoStop}%
\bibitem [{\citenamefont {Tucker}\ \emph {et~al.}(2012)\citenamefont {Tucker},
  \citenamefont {Fernandes}, \citenamefont {Li}, \citenamefont {Thampy},
  \citenamefont {Ni}, \citenamefont {Abernathy}, \citenamefont {Bud'ko},
  \citenamefont {Canfield}, \citenamefont {Vaknin}, \citenamefont {Schmalian},\
  and\ \citenamefont {McQueeney}}]{mcqueeneyprb12}%
  \BibitemOpen
  \bibfield  {author} {\bibinfo {author} {\bibfnamefont {G.~S.}\ \bibnamefont
  {Tucker}}, \bibinfo {author} {\bibfnamefont {R.~M.}\ \bibnamefont
  {Fernandes}}, \bibinfo {author} {\bibfnamefont {H.~F.}\ \bibnamefont {Li}},
  \bibinfo {author} {\bibfnamefont {V.}~\bibnamefont {Thampy}}, \bibinfo
  {author} {\bibfnamefont {N.}~\bibnamefont {Ni}}, \bibinfo {author}
  {\bibfnamefont {D.~L.}\ \bibnamefont {Abernathy}}, \bibinfo {author}
  {\bibfnamefont {S.~L.}\ \bibnamefont {Bud'ko}}, \bibinfo {author}
  {\bibfnamefont {P.~C.}\ \bibnamefont {Canfield}}, \bibinfo {author}
  {\bibfnamefont {D.}~\bibnamefont {Vaknin}}, \bibinfo {author} {\bibfnamefont
  {J.}~\bibnamefont {Schmalian}}, \ and\ \bibinfo {author} {\bibfnamefont
  {R.~J.}\ \bibnamefont {McQueeney}},\ }\href@noop {} {\bibfield  {journal}
  {\bibinfo  {journal} {Phys. Rev. B}\ }\textbf {\bibinfo {volume} {86}},\
  \bibinfo {pages} {024505} (\bibinfo {year} {2012})}\BibitemShut {NoStop}%
\bibitem [{\citenamefont {Qi}\ \emph {et~al.}()\citenamefont {Qi},
  \citenamefont {Raghu}, \citenamefont {Liu}, \citenamefont {Scalapino},\ and\
  \citenamefont {Zhang}}]{scalapinoarXiv08}%
  \BibitemOpen
  \bibfield  {author} {\bibinfo {author} {\bibfnamefont {X.-L.}\ \bibnamefont
  {Qi}}, \bibinfo {author} {\bibfnamefont {S.}~\bibnamefont {Raghu}}, \bibinfo
  {author} {\bibfnamefont {C.-X.}\ \bibnamefont {Liu}}, \bibinfo {author}
  {\bibfnamefont {D.~J.}\ \bibnamefont {Scalapino}}, \ and\ \bibinfo {author}
  {\bibfnamefont {S.-C.}\ \bibnamefont {Zhang}},\ }\href@noop {} {}\bibinfo
  {howpublished} {arXiv:0804.4332}\BibitemShut {NoStop}%
\bibitem [{\citenamefont {Ran}\ \emph {et~al.}(2009)\citenamefont {Ran},
  \citenamefont {Wang}, \citenamefont {Zhai}, \citenamefont {Vishwanath},\ and\
  \citenamefont {Lee}}]{dunghailee09}%
  \BibitemOpen
  \bibfield  {author} {\bibinfo {author} {\bibfnamefont {Y.}~\bibnamefont
  {Ran}}, \bibinfo {author} {\bibfnamefont {F.}~\bibnamefont {Wang}}, \bibinfo
  {author} {\bibfnamefont {H.}~\bibnamefont {Zhai}}, \bibinfo {author}
  {\bibfnamefont {A.}~\bibnamefont {Vishwanath}}, \ and\ \bibinfo {author}
  {\bibfnamefont {D.-H.}\ \bibnamefont {Lee}},\ }\href {\doibase
  10.1103/PhysRevB.79.014505} {\bibfield  {journal} {\bibinfo  {journal} {Phys.
  Rev. B}\ }\textbf {\bibinfo {volume} {79}},\ \bibinfo {pages} {014505}
  (\bibinfo {year} {2009})}\BibitemShut {NoStop}%
\bibitem [{\citenamefont {Lau}\ and\ \citenamefont
  {Timm}(2013)}]{lautimmprb13}%
  \BibitemOpen
  \bibfield  {author} {\bibinfo {author} {\bibfnamefont {A.}~\bibnamefont
  {Lau}}\ and\ \bibinfo {author} {\bibfnamefont {C.}~\bibnamefont {Timm}},\
  }\href@noop {} {\bibfield  {journal} {\bibinfo  {journal} {Phys. Rev. B}\
  }\textbf {\bibinfo {volume} {88}},\ \bibinfo {pages} {165402} (\bibinfo
  {year} {2013})}\BibitemShut {NoStop}%
\bibitem [{\citenamefont {Castellani}\ \emph {et~al.}(1978)\citenamefont
  {Castellani}, \citenamefont {Natoli},\ and\ \citenamefont
  {Ranninger}}]{castellani78}%
  \BibitemOpen
  \bibfield  {author} {\bibinfo {author} {\bibfnamefont {C.}~\bibnamefont
  {Castellani}}, \bibinfo {author} {\bibfnamefont {C.~R.}\ \bibnamefont
  {Natoli}}, \ and\ \bibinfo {author} {\bibfnamefont {J.}~\bibnamefont
  {Ranninger}},\ }\href {\doibase 10.1103/PhysRevB.18.4945} {\bibfield
  {journal} {\bibinfo  {journal} {Phys. Rev. B}\ }\textbf {\bibinfo {volume}
  {18}},\ \bibinfo {pages} {4945} (\bibinfo {year} {1978})}\BibitemShut
  {NoStop}%
\bibitem [{\citenamefont {Fanfarillo}\ \emph {et~al.}()\citenamefont
  {Fanfarillo}, \citenamefont {Cortijo},\ and\ \citenamefont
  {Valenzuela}}]{workinprogress}%
  \BibitemOpen
  \bibfield  {author} {\bibinfo {author} {\bibfnamefont {L.}~\bibnamefont
  {Fanfarillo}}, \bibinfo {author} {\bibfnamefont {A.}~\bibnamefont {Cortijo}},
  \ and\ \bibinfo {author} {\bibfnamefont {B.}~\bibnamefont {Valenzuela}},\
  }\href@noop {} {}\bibinfo {howpublished} {Work in progress}\BibitemShut
  {NoStop}%
\bibitem [{\citenamefont {Avci}\ \emph {et~al.}(2014)\citenamefont {Avci},
  \citenamefont {Chmaissem}, \citenamefont {Allred}, \citenamefont
  {Rosenkranz}, \citenamefont {Eremin}, \citenamefont {Chubukov}, \citenamefont
  {Bugaris}, \citenamefont {Chung}, \citenamefont {Kanatzidis}, \citenamefont
  {Castellan}, \citenamefont {Schlueter}, \citenamefont {Claus}, \citenamefont
  {Khalyavin}, \citenamefont {Manuel}, \citenamefont {Daoud-Aladine},\ and\
  \citenamefont {Osborn}}]{nematicnatcomm14}%
  \BibitemOpen
  \bibfield  {author} {\bibinfo {author} {\bibfnamefont {S.}~\bibnamefont
  {Avci}}, \bibinfo {author} {\bibfnamefont {O.}~\bibnamefont {Chmaissem}},
  \bibinfo {author} {\bibfnamefont {J.}~\bibnamefont {Allred}}, \bibinfo
  {author} {\bibfnamefont {S.}~\bibnamefont {Rosenkranz}}, \bibinfo {author}
  {\bibfnamefont {I.}~\bibnamefont {Eremin}}, \bibinfo {author} {\bibfnamefont
  {A.}~\bibnamefont {Chubukov}}, \bibinfo {author} {\bibfnamefont
  {D.}~\bibnamefont {Bugaris}}, \bibinfo {author} {\bibfnamefont
  {D.}~\bibnamefont {Chung}}, \bibinfo {author} {\bibfnamefont
  {M.}~\bibnamefont {Kanatzidis}}, \bibinfo {author} {\bibfnamefont {J.-P.}\
  \bibnamefont {Castellan}}, \bibinfo {author} {\bibfnamefont {J.}~\bibnamefont
  {Schlueter}}, \bibinfo {author} {\bibfnamefont {H.}~\bibnamefont {Claus}},
  \bibinfo {author} {\bibfnamefont {D.}~\bibnamefont {Khalyavin}}, \bibinfo
  {author} {\bibfnamefont {P.}~\bibnamefont {Manuel}}, \bibinfo {author}
  {\bibfnamefont {A.}~\bibnamefont {Daoud-Aladine}}, \ and\ \bibinfo {author}
  {\bibfnamefont {R.}~\bibnamefont {Osborn}},\ }\href@noop {} {\bibfield
  {journal} {\bibinfo  {journal} {Nature Communications}\ }\textbf {\bibinfo
  {volume} {5}} (\bibinfo {year} {2014})}\BibitemShut {NoStop}%
\bibitem [{\citenamefont {Calderon}\ \emph {et~al.}(2009)\citenamefont
  {Calderon}, \citenamefont {Valenzuela},\ and\ \citenamefont
  {Bascones}}]{nosotrasnjp09}%
  \BibitemOpen
  \bibfield  {author} {\bibinfo {author} {\bibfnamefont {M.~J.}\ \bibnamefont
  {Calderon}}, \bibinfo {author} {\bibfnamefont {B.}~\bibnamefont
  {Valenzuela}}, \ and\ \bibinfo {author} {\bibfnamefont {E.}~\bibnamefont
  {Bascones}},\ }\href@noop {} {\bibfield  {journal} {\bibinfo  {journal} {New
  J. Phys.}\ }\textbf {\bibinfo {volume} {11}},\ \bibinfo {pages} {013051}
  (\bibinfo {year} {2009})}\BibitemShut {NoStop}%
\bibitem [{\citenamefont {Song}\ \emph
  {et~al.}(2013{\natexlab{b}})\citenamefont {Song}, \citenamefont {Liang},
  \citenamefont {Lim},\ and\ \citenamefont {Haas}}]{stephanhaasprb13}%
  \BibitemOpen
  \bibfield  {author} {\bibinfo {author} {\bibfnamefont {K.~W.}\ \bibnamefont
  {Song}}, \bibinfo {author} {\bibfnamefont {Y.-C.}\ \bibnamefont {Liang}},
  \bibinfo {author} {\bibfnamefont {H.}~\bibnamefont {Lim}}, \ and\ \bibinfo
  {author} {\bibfnamefont {S.}~\bibnamefont {Haas}},\ }\href@noop {} {\bibfield
   {journal} {\bibinfo  {journal} {Phys. Rev. B}\ }\textbf {\bibinfo {volume}
  {88}},\ \bibinfo {pages} {054501} (\bibinfo {year}
  {2013}{\natexlab{b}})}\BibitemShut {NoStop}%
\bibitem [{\citenamefont {Negele}\ and\ \citenamefont {Orland}(1988)}]{negele}%
  \BibitemOpen
  \bibfield  {author} {\bibinfo {author} {\bibfnamefont {J.}~\bibnamefont
  {Negele}}\ and\ \bibinfo {author} {\bibfnamefont {H.}~\bibnamefont
  {Orland}},\ }\href@noop {} {\emph {\bibinfo {title} {Quantum many particle
  system}}}\ (\bibinfo  {publisher} {Addison-Wesley, New York},\ \bibinfo
  {year} {1988})\BibitemShut {NoStop}%
\bibitem [{\citenamefont {Benfatto}\ \emph {et~al.}(2004)\citenamefont
  {Benfatto}, \citenamefont {Toschi},\ and\ \citenamefont
  {Caprara}}]{benfatto04}%
  \BibitemOpen
  \bibfield  {author} {\bibinfo {author} {\bibfnamefont {L.}~\bibnamefont
  {Benfatto}}, \bibinfo {author} {\bibfnamefont {A.}~\bibnamefont {Toschi}}, \
  and\ \bibinfo {author} {\bibfnamefont {S.}~\bibnamefont {Caprara}},\
  }\href@noop {} {\bibfield  {journal} {\bibinfo  {journal} {Phys. Rev. B}\
  }\textbf {\bibinfo {volume} {69}},\ \bibinfo {pages} {184510} (\bibinfo
  {year} {2004})}\BibitemShut {NoStop}%
\bibitem [{\citenamefont {Fanfarillo}\ \emph {et~al.}(2009)\citenamefont
  {Fanfarillo}, \citenamefont {Benfatto}, \citenamefont {Caprara},
  \citenamefont {Castellani},\ and\ \citenamefont {Grilli}}]{laura09}%
  \BibitemOpen
  \bibfield  {author} {\bibinfo {author} {\bibfnamefont {L.}~\bibnamefont
  {Fanfarillo}}, \bibinfo {author} {\bibfnamefont {L.}~\bibnamefont
  {Benfatto}}, \bibinfo {author} {\bibfnamefont {S.}~\bibnamefont {Caprara}},
  \bibinfo {author} {\bibfnamefont {C.}~\bibnamefont {Castellani}}, \ and\
  \bibinfo {author} {\bibfnamefont {M.}~\bibnamefont {Grilli}},\ }\href@noop {}
  {\bibfield  {journal} {\bibinfo  {journal} {Phys. Rev. B}\ }\textbf {\bibinfo
  {volume} {79}},\ \bibinfo {pages} {172508} (\bibinfo {year}
  {2009})}\BibitemShut {NoStop}%
\bibitem [{\citenamefont {Marciani}\ \emph {et~al.}(2013)\citenamefont
  {Marciani}, \citenamefont {Fanfarillo}, \citenamefont {Castellani},\ and\
  \citenamefont {Benfatto}}]{laura13}%
  \BibitemOpen
  \bibfield  {author} {\bibinfo {author} {\bibfnamefont {M.}~\bibnamefont
  {Marciani}}, \bibinfo {author} {\bibfnamefont {L.}~\bibnamefont
  {Fanfarillo}}, \bibinfo {author} {\bibfnamefont {C.}~\bibnamefont
  {Castellani}}, \ and\ \bibinfo {author} {\bibfnamefont {L.}~\bibnamefont
  {Benfatto}},\ }\href@noop {} {\bibfield  {journal} {\bibinfo  {journal}
  {Phys. Rev. B}\ }\textbf {\bibinfo {volume} {88}},\ \bibinfo {pages} {214508}
  (\bibinfo {year} {2013})}\BibitemShut {NoStop}%
\bibitem [{not()}]{note_orbmagn}%
  \BibitemOpen
  \href@noop {} {\bibinfo  {journal} {This is a direct consequence of
  neglecting interorbital magnetization in the original Hamiltonian i.e. of
  considering in the spin-operator $\vec{S}_\eta$ definition only fermions
  coming from the same orbital $\eta$}\ }\BibitemShut {NoStop}%
\end{thebibliography}%

\end{document}